\documentclass[pdflatex,sn-mathphys-num]{sn-jnl}% Math and Physical Sciences Numbered Reference Style
\usepackage{graphicx}%
\usepackage{multirow}%
\usepackage{amsmath,amssymb,amsfonts}%
\usepackage{amsthm}%
\usepackage{mathrsfs}%
\usepackage[title]{appendix}%
\usepackage{xcolor}%
\usepackage{textcomp}%
\usepackage{manyfoot}%
\usepackage{booktabs,tabularx,makecell}%
\usepackage{algorithm}%
\usepackage{algorithmicx}%
\usepackage{algpseudocode}%
\usepackage{listings}%
\usepackage{multicol}%
\usepackage{paracol}%

\usepackage{amsmath,amssymb,mathtools}
\usepackage{stmaryrd}   % for \llbracket \rrbracket
\usepackage{mathpartir} % for \inferrule

\newcommand{\toggleBlock}[1]{}

\newcommand{\teq}{\mathrel{\preccurlyeq}}               % total order on ties

\lstdefinestyle{EBNF}{%
  frame=l, numbers=left, numbersep=1em, xleftmargin=2em, 
  basicstyle=\ttfamily \tiny,
  moredelim=[is][\underline]{'}{'},
  breaklines=true,
}%

\lstdefinestyle{CodeBlock}{%
  frame=l, numbers=left, numbersep=1em, xleftmargin=2em, 
  basicstyle=\ttfamily \tiny,
  breaklines=true,
}%

\lstdefinestyle{nbr}{%
    numbers=left, numbersep=1em,
}%

\theoremstyle{thmstyleone}%
\theoremstyle{thmstyletwo}%

\theoremstyle{thmstylethree}%

\begin{document}

\title[M]{M, Toolchain and Language for Reusable Model Compilation}


\author*[1]{\fnm{Hiep} \sur{Hong Trinh}}\email{hiep.hong.trinh@mdu.se}

\author[1]{\fnm{Federico} \sur{Ciccozzi}}\email{federico.ciccozzi@mdu.se}
%\equalcont{}

\author[1]{\fnm{Abu Naser} \sur{Masud}}\email{abu.naser.masud@mdu.se}

\author[1]{\fnm{Marjan} \sur{Sirjani}}\email{marjan.sirjani@mdu.se}
%\equalcont{}

\author[1]{\fnm{Mikael} \sur{Sjödin}}\email{mikael.sjodin@mdu.se}

\affil[1]{\orgdiv{IDT school}, \orgname{Mälardalen University}, \orgaddress{\city{Västerås}, \country{Sweden}}}


\abstract{
Complex software-driven systems—such as cyber-physical systems (CPS), Internet of Things (IoT) platforms, multi-agent systems (MAS), and autonomous robotics—often combine distributed concurrent computation with continuous interaction with the physical environment. Engineering them benefits from actionable software models systematically transformed into specialized artifacts for simulation, verification, or execution, often requiring distinct formalisms, specification languages, or platforms.

Traditional compilers translate programs into executables. By analogy, a model compiler should take a system model and generate diverse target artifacts while preserving semantics. However, most modeling languages are designed for a single target, typically simulation or implementation, and lack integrated support for multi-target compilation. This creates inefficiencies and hinders the transition from abstract design to practical deployment and verification.

We present M, a grammar-driven textual modeling language and toolchain that addresses this limitation. M is based on the actor model and extended with discrete-event scheduling semantics, enabling intuitive specification of system entities, their message-based interactions, and their time- or state-dependent behaviors. M also functions as a middle language; existing modeling languages can be mapped to M to reuse its multi-target compilation capabilities. The toolchain ensures semantic consistency between generated artifacts, allowing seamless movement between simulation, execution, and verification.

By integrating modeling, compilation, and semantic conformance, M advances model-driven engineering for complex concurrent timed systems. This approach has the potential to reduce development effort, increase reliability, and promote reuse across domains where concurrency and timing are central.
}




\maketitle

\section{Introduction} \label{sec:intro}
The development of complex cyber-physical systems (CPS), Internet-of-Things (IoT) infrastructures, and robotics platforms increasingly relies on precise models that capture both discrete and continuous aspects of system behavior. Such models serve multiple purposes: they are used for early simulation and design-space exploration, for deployment to real platforms, and for rigorous analysis through formal verification. However, existing modeling languages and toolchains often emphasize one dimension—either expressiveness for simulation, executability for deployment, or analyzability for verification—while providing limited support for seamless transition across these domains.

To address this gap, we introduce M, a modeling language designed with semantics-aware compilation as a central principle. M combines the actor-based concurrency model with discrete event (DE) scheduling and explicit support for temporal semantics, enabling the representation of timeline and state variables as time series. This foundation not only accommodates reactive and distributed architectures commonly found in CPS and IoT systems, but also extends naturally to continuous dynamics by integrating discretization mechanisms. Consequently, M provides a unified platform where models can be systematically transformed into code for simulation, verification, or real deployment without manual intervention.

The key insight behind M is that model compilation should preserve semantics across abstraction levels. Rather than treating an executable as the final endpoint of modeling, M positions executable artifacts as one of several possible derivations from a high-level description. By defining precise operational semantics and a structured compilation pipeline, M enables the reuse of the same model for multiple purposes, thereby reducing redundancy and mitigating the risks of semantic drift. Moreover, by adopting a textual language design grounded in grammar-based syntax, M benefits from well-established compiler theory while enabling integration with modern language technologies, including automated translation and synthesis.

This paper presents the conceptual foundations and design of the M language. We motivate its development by analyzing limitations of current modeling approaches, describing its syntax and semantics, and demonstrating how M enables reusable model compilation for heterogeneous targets. Through this work, we aim to contribute not only a practical language but also a broader methodology for bridging modeling, simulation, deployment, and formal verification in a unified framework.

The origins of M trace back to a master’s thesis project \cite{hiepmasterthesis}, later extended into research papers \cite{HongTrinh6940, ros2rebeca_journal}. In that work, we addressed the challenge of modeling and verifying safety properties of a multi-robot system in ROS2\footnote{\url{https://ros.org/}}
 using the Timed Rebeca\footnote{\url{https://rebeca-lang.org/}} language \cite{DBLP:journals/scp/ReynissonSACJIS14}
 and its model checker. The actor model \cite{Agha1986Actors} and class-based structure of Rebeca provided a natural way to capture reactive node graphs, as commonly encountered in ROS2, CPS, and IoT systems. Moreover, Timed Rebeca’s explicit notion of time enabled the modeling of time-constrained and time-convertible behaviors.

However, the experience also revealed limitations: the discrete abstraction of inherently continuous behaviors, while necessary for tractable model checking, often introduced approximation errors and semantic misalignments between the model and the real system. These difficulties motivated us to envision a more general modeling language—one not tied to model checking as its primary goal, but capable of supporting both discrete and continuous dynamics, while also enabling code generation for simulation, deployment, and verification. Importantly, such a language should shield users from the complexity of mapping mechanisms and correctness checks, allowing them to focus on high-level abstractions of the system under study.

This vision aligned with the objectives of the \href{https://www.es.mdu.se/projects/603-ORPHEUS__systematic_and_reliable_theory_for_model_compilation}{ORPHEUS} project, which called for the development of precise theories and prototype demonstrations of semantics-aware model compilation. In response, we designed an end-to-end toolchain (see Fig.~\ref{fig:m}) and systematically investigated candidate languages to serve as its core. Ultimately, after evaluating the latest solutions, we concluded that a new modeling language was necessary, leading to the creation of M as the foundation of the toolchain.

The remainder of this paper is structured as follows. Section \ref{sec:lit} walks through the state-of-the-art. Section \ref{sec:toolchain} introduces the overall M toolchain. Section \ref{sec:design} details the design of the M language, including its formal semantics, language constructs, syntax, and levels of expressiveness. Section \ref{sec:types} focuses on the type system and typing rules, which is an essential part of an executable semantics. Section \ref{sec:trans} demonstrates M’s transformability into other formalisms and languages. Section \ref{sec:cmp} provides a comparative analysis with existing approaches. Section \ref{sec:progress} outlines current progress and future development plans. Finally, Section \ref{sec:conclude} concludes the paper.
%The remainder of the paper is organized as follows. Section \ref{sec:lit} walks through state of the art. Section \ref{sec:toolchain} presents the overall M toolchain. Section \ref{sec:design} presents the design of our M language candidate, its semantics, language constructs and syntax, high-level and low-level expressiveness. Section \ref{sec:types} is dedicated to the type system and typing rules of M which is an essential part of an executable semantics. Section \ref{sec:trans} shows M's transformability to other formalisms and languages. Section \ref{sec:cmp} compares it with several existing languages. Section \ref{sec:progress} is about our current progress and future release plans. Finally, section \ref{sec:conclude} concludes the paper.

\section{Background and State-of-art Review} \label{sec:lit}
In this section, we survey background and state of the art: it frames software-based models as first-class artifacts for simulation, deployment, and formal verification, motivates textual languages as a compilation-ready substrate, and reviews prevailing languages and tools to expose semantic discontinuities that motivate a semantics-aware compilation pipeline.
\subsection{A View of Software-based Models}
In systems and software engineering, we consider \emph{software-based models}, understood as dynamically executable computerized models. Such models can be executed in different modes:
\begin{itemize}
    \item \textbf{Simulation:} the model is executed in a virtual environment along an artificial 
    (logical) timeline.
    \item \textbf{Real deployment:} the model is implemented, deployed, and executed in the real 
    environment along the physical timeline.
    \item \textbf{Formal verification:} most notably through model checking, where the model is 
    systematically explored across all possible scenarios to verify that required safety properties 
    hold in every case.
\end{itemize}
From this perspective, we treat software-based models as first-class software artifacts, engineered using similar methods and tooling as conventional software: programming languages, compilers, IDEs, and execution platforms. Graphical models ultimately require storage and reconstruction in a textual format, such as XML, JSON, or grammar-conformant text. Therefore, textual modeling languages are particularly relevant for compilation: grammar-driven texts can be precisely specified, parsed, and compiled using well-established principles of programming language and compiler theory 
\cite{10.1145/568438.568455,10.1145/230514.571645,AhoLSU2006}. 
The emergence of large language models (LLMs) further reinforces this relevance, as they enable novel approaches to language-to-language translation.

Each model is a purposeful, non-comprehensive representation of the real system. No universal model exists beyond the system itself, which is typically too complex and costly for direct analysis or experimentation. Models are not self-contained artifacts: they inevitably need to interoperate with other levels of abstraction and alternative representations. Historically, compiler theory has treated executable code as the final destination of the compilation process. In modeling, however, the executable is better regarded as a \emph{terminal artifact}, where the transformation chain stops. Reverse engineering from executables to recover high-level models is generally infeasible. Executables should instead be understood as the lowest abstraction level consumable by a given runtime environment: for example, a web server executes HTML and JavaScript, a simulator compiles and executes its surface language, and a compiler builds applications from a codebase. To derive 
these artifacts from higher-level models, one follows a series of lowering steps, ultimately reaching a format accepted by the target compiler, generator, or executor.

At each abstraction level, a \emph{semantic discontinuity} \cite{Bran94} may arise, posing risks of semantic loss or discrepancy in translation. Preserving semantics is therefore essential to validate equivalence between the source model and its derived representations. Modeling languages—especially graphical ones such as UML—have often been criticized for lacking a well-defined semantic foundation, leading to multiple interpretations of modeling constructs \cite{alh2012uml,lund2010}. Behavioural logic is frequently externalized through an action language (e.g., Alf for UML \cite{umlalf}) or embedded as inline user code (e.g., ``opaque actions'' in SysML 
v2 \cite{DBLP:journals/jot/JansenPRSW22}). Such practices introduce semantic obscurity and further widen semantic gaps during translation.

Many languages, frameworks, and platforms provide effective support for specific stages of model-based development. However, a comprehensive mechanism for adapting and integrating these components—enabling seamless end-to-end development and maximizing reusability—remains absent.

\subsection{The Landscape of Modeling Languages and Tools}
In this section, we do not attempt to systematically and exhaustively classify all existing modeling languages and tools. Instead, we briefly discuss the main modeling approaches used by different research and engineering communities to identify their common needs and challenges and find a way to connect their best practices. Many parts can be found in the Fishwick handbook~\cite{Fishwick2007}.

In software engineering, UML and MOF/MDA-style metamodeling are widespread. The OMG stack distinguishes four abstraction levels (M0-M3), and model transformations, model-to-model (M2M) and model-to-text (M2T), are defined against metamodels. These metamodels mainly capture structure; precise operational semantics is usually introduced later via action languages or profiles rather than being intrinsic to the metamodel itself (e.g., foundational UML/Alf~\cite{umlalf}). Executability is therefore not the default in practice, and tool providers often rely on 'opaque' behavioral hooks that blur semantics (e.g., in SysML v2). This combination explains why seemingly similar diagrams may admit multiple, tool-specific interpretations and why transformations risk semantic drift.

The formal verification community relies on mathematically rigorous models that enable algorithmic analysis of correctness properties. Typical mathematical objects in use are abstract machines (finite-state automata, labeled transition systems), timed abstract machines (timed automata \cite{alur1994theory}, hybrid automata \cite{henzinger1996the}), concurrency models (Hoare logic, Petri net, process algebras \cite{baeten2005process}), and formal semantics definitions including operational, denotational, and axiomatic semantics (see e.g. \cite{winskel1993semantics}). Verification-oriented models, however, often adopt a significant amount of simplifications (abstractions, idealized timing, state space limitations) to ensure decidability and tractability, which can hinder direct application to large, intricate systems. At the same time, the strict formalism is the key strength of this community. Our stance is that, by relaxing proof obligations while preserving precise operational semantics, verification formalisms can be repurposed as robust executable models for more general modeling and simulation. The formalism/semantics-first design of M aligns with this observation.

The simulation modeling community is familiar with the use of differential equations to simulate continuous systems and discrete events for discrete systems, as described by Zeigler et al.~\cite{devs}, Banks \cite{banks2010discrete}, Law \cite{law2014simulation}, and Wainer \cite{devs2}. They model systems with a clear purpose: to simulate the happenings in the system chronologically using a computerized program. In DE simulation, the system changes state only at discrete instants when events occur. Event times may be deterministic or sampled from specified probability distributions. A time-ordered future-event list (FEL) maintains pending events; after each event is processed, the FEL is updated to reflect the new state—cancelling invalidated events and inserting any newly scheduled ones with their timestamps. A simulation is done as a walk through an artificial history of system state. Starting at time 0, the simulator repeatedly removes the imminent event - the one with the smallest timestamp - from the future event list, advances the logical clock to that time, executes the event's effects, and continues. DEVS, Parallel DEVS, and other variants are popular formalisms. There are DEVS or Parallel DEVS-compliant simulators that can run simulation models written in a programming language or a dedicated description language. The simulation-modeling community follows a well-established process \cite{banks2010discrete} in which model verification and validation are explicit, integral activities.

%Hiep's original text: The simulation modeling community is familiar with using differential equations for simulating continuous systems (DESS) and discrete events for discrete systems (DES). Best literature are books written by Bernard Zeigler et al.\cite{devs}, Jerry Banks\cite{banks2010discrete}, Averill M.Law\cite{law2014simulation}, and Gabriel Wainer\cite{devs2}. They model systems with a clear purpose: to simulate the happenings in the system chronically using a computerized program. In DES, the progression of a system's state is seen as a sequence of predictable events whose stochastic occurrences can be modelled using different patterns of probabilistic distribution. A future event list (FEL) dynamically keeps track of upcoming events - after each event, up to the new situation, previously scheduled events may be removed from or new events may be added to the list. A simulation is done as a walk through an artificial history of system state, starting at time zero then the logical clock is advanced artificially by the next-event time, i.e the event with the smallest timestamp taken out of the future event list. DEVS, Parallel DEVS and other variants are popular formalisms. There are DEVS or Parallel DEVS-compliant simulators that can run simulation models written in a programming language or dedicated description language. The simulation modeling community has a clear modeling process\cite[p.15]{banks2010discrete} that addresses model verification and validation explicitly.%

The MBSE community routinely combines multiple languages and tools for modeling, designing, simulating, and creating complex systems that include both physical and software components, e.g., SysML/UML for requirements/architecture, Modelica for physical dynamics, AADL for platform/architecture analysis, and MATLAB/Simulink for control logic and signal processing (see e.g. \cite{sae2018sysml,sae2014codegen,vallecillo2021systemlevel}). These workflows must capture both physical and software behavior; accordingly, they rely on the model's executability for simulation, automation of code generation for software components and verification of system-level properties across artifacts.

%The MBSE community routinely composes multiple languages and tools—SysML/UML for requirements and architecture, Modelica for physical dynamics, AADL for platform/architecture analysis, and MATLAB/Simulink for control and signal processing—to model and simulate complex cyber–physical systems \cite{}. These workflows must capture both software and physical behavior; accordingly, they rely on executable models, automatic code generation for software components, and verification of system-level properties across artifacts \cite{}.

    %\item F2: Languages based on formal foundation of textual grammar-conformant languages with a precise executable semantics: Modelica, ThingML, Lingua Franca, ABS.
    %\item F3: Languages based on data description (E/R diagram, XML, JSON). These languages are used to define data schemas, structures, and serialization formats, often without behavior. The semantics is not embedded in the language but depends on how the data consumer application interpretes.    

\subsection{Problems and Challenges}
Selic’s landmark review and vision paper \cite{DBLP:journals/sosym/Selic12} remains strikingly relevant more than a decade later. Despite repeated promises of higher abstraction and automation, models still do not occupy the central and authoritative role in software design that they do in traditional engineering disciplines. The design of modeling languages often lacks a systematic engineering discipline and a mature theoretical foundation comparable to that of programming languages. In particular, semantics definition remains a persistent challenge: it should be addressed through principled combinations of higher-level and lower-level formalisms. Complex systems typically require multiple models across different languages; these must be synchronized to preserve consistency, yet interchange mechanisms between tools are frequently missing or inadequate. Model transformation is likewise not an established, repeatable engineering task; it needs a widely agreed theoretical underpinning, one that can profitably reuse results from compiler theory. The resulting semantic discontinuity between source models and executable artifacts is a major source of risk, and without a formal foundation for modeling languages, it is difficult to argue that critical requirements are, in fact, met.

Building on this diagnosis, Bo Liu et al. \cite{DBLP:journals/jzusc/LiuZCLGW20} survey model-driven techniques and tools for cyber-physical systems and foresee the need for an integrated, highly scalable platform with a unified modeling language of strict syntax and rich semantics, acting as a bridge to other model-driven tools via bidirectional adapters. Among the works that point in this direction is Karma \cite{karma1,karma2,karma3}, which leverages the EMF Ecore metamodel as a common foundation. Complementarily, Lano et al. \cite{DBLP:journals/jss/LanoRTS18} catalogue design patterns for model transformation; notably, introducing an intermediate metamodel reduces the transformation burden from \textsl{N×M} pairwise mappings to \textsl{N+M} by adapting sources to, and targets from, the intermediate layer.

Conversions between modeling languages are ubiquitous: textualizing graphical notations such as UML \cite{textuml}; generating text from metamodels \cite{textmm,textmm2}; producing formal verification models from UML \cite{verifyuml}, SysML \cite{DBLP:journals/jacic/SaquiSannesAV21} or Lingua Franca \cite{lf2rebeca}; and mapping across languages (e.g., Modelica to UML \cite{schamai2009modelica}). A wide variety of model-to-model transformation languages and patterns are documented in \cite{DBLP:journals/jss/LanoRTS18}.

Within the simulation community, particularly DE simulation, there have been several efforts to define high-level specification layers over DEVS \cite{devs,devs2} and its variants—via XML (DEVSML \cite{devsml1,devsml2}), MDE metamodels, E/R diagrams (surveyed in \cite{7095017}), and context-free languages (CFG\_DEVS \cite{10.5555/3643142.3643356}, DEVSpecL \cite{devspecl}, CML-DEVS \cite{cmldevs1,cmldevs2}). There are also numerous attempts to integrate simulation with formal verification by translating constrained subsets of DEVS into Timed Automata and related formalisms for model checking \cite{DBLP:journals/ijspm/ZeiglerNS17}.

Taken together, these observations reinforce a well-established conclusion: no single model suffices for the design, construction, and analysis of complex systems. What is needed is an integrated platform that connects multiple purpose-specific models through disciplined adapters and transformations, organized around a central model that captures essential structure and behavior while preserving semantics across the toolchain.

M operationalizes this vision by serving as a semantics-aware pivot that concentrates meaning in a single executable model while allowing diverse artifacts to connect at the edges. Concretely, models from other DSLs are adapted once into M - optionally via intermediate textualizations - so that parsing and semantic analysis happen a single time;  multiple generators can then reuse the resulting decorated AST. This preserves a common interpretation of structure and behavior while fanning out to targets such as simulation artifacts, inputs to domain-specific compilers and code generators, direct executables through the MLIR/LLVM stack, and even verification models (e.g., Timed Rebeca/UPPAAL). The approach turns the classic \textsl{N×M} mapping problem into \textsl{N+M}: \textsl{N} adapters from source notations into \textsl{M}, and \textsl{M} generators from M to platforms. Crucially, M's actor-based, DE operational semantics (messages, queues, time advance, scheduling/cancellation) make the central model both precise and directly executable, which helps maintain semantic consistency across the toolchain as artifacts are produced for different purposes.

\subsection{The Actor Model and Actor Languages}
Our M language is inspired by Timed Rebeca, the pioneer in a family of actor-based languages like Lingua Franca\footnote{\url{https://www.lf-lang.org/}}, P\footnote{\url{https://p-org.github.io/P/}}, ABS\footnote{\url{https://abs-models.org/}}.
% Therefore, a brief understanding of how the actor model works is necessary for self-inclusiveness of the paper.
 The actor model, first formulated by Hewitt in 1973 \cite{DBLP:conf/ijcai/HewittBS73} then revamped by Gul in 1986 \cite{Agha1986Actors}, provides a simple yet robust architecture to reason about the liveness of concurrent, asynchronous systems. The essence of the classic actor model lies in the following principles:
\begin{itemize}
    \item Unshared state: Each actor is an independent processing unit with its local state shielded from other actors. Actors can only interact by sending asynchronous messages to each other to tell them what to do.
    \item Non-blocked message passing: the sender just sends without waiting for the receiver's readiness or response, just like a man drops a mail to a postbox then goes away doing other jobs. Messages arrive at the receiver's side in a so-called mailbox, which can be a sorted queue or unsorted bag. An actor also can send messages to itself. 
    \item Uninterrupted sequential task processing: An actor processes messages in a sequential, run-to-completion style - one by one, no blocking, no interruption until finished. It is up to the receiver to decide which message to pick up at a time, normally first-come-first-serve, and only one is picked each time. This principle is referred to as 'isolated turn principle' by Koster in \cite{10.1145/3001886.3001890}.
\end{itemize}
Because each actor processes tasks sequentially, there is no local data racing since its local state is only assessed by the one active task at a time. The mechanisms of non-blocked sending, local queuing, and sequential processing of messages allow robust concurrency and liveness. The actor model helps relieve users from complicated mutual access synchronization mechanisms, typically found in multi-threaded programming, to focus only on interaction flows in the system.

\subsection{DEVS and Parallel DEVS}
%\subsection{Discrete Event System Specification (DEVS) and Parallel DEVS}
As M integrates the DE semantics into the actor model, we briefly introduce the DEVS (Discrete Event System Specification) formalism, which was formulated by Zeigler \cite{zeigler1976theory} in the 1970s and is still widely used in the simulation modeling community. DEVS is a modular and hierarchical framework for modeling DE dynamic systems. In DEVS, system structure and behaviour are expressed as a set of coupled and atomic models. An atomic DEVS model has the form: \[A=\langle s \in S,ta:S \to \Re_{\geq 0},\delta_{int}:S \to S,X,\delta_{ext}:S\times\Re_{\geq 0}\times X \to S,Y,\lambda:S\to Y \rangle\] where:
\begin{itemize}
    \item S is the state space, which can be infinite, X is the input set, and Y is the output set which contains $\varnothing$, meaning no output is generated,
    \item $s\in S$ is its internal state, whose lifetime is decided by the time advance function $ta(s)$,
    \item At the end of the current state's lifetime, an output function $\lambda$ is activated to generate output $y=\lambda(s) \in Y$, then the state is switched to $s'=\delta_{int}(s)$ with a lifetime of $ta(s')$,
    \item During the lifetime of a state $s$, an input signal $x\in X$ may arrive and trigger the state to switch to $s'=\delta_{ext}(s,e,x)$ where $0\leq e \leq ta(s)$ is the elapsed state time. Likewise, the lifetime of the new state is reset to $ta(s')$.
\end{itemize}
Coupled models are compositions of atomic or other coupled models using the input-output coupling, enabling hierarchical system construction. Parallel DEVS or PDEVS \cite{717419,TendelooV18} extends the classic DEVS formalism by allowing multiple concurrent input signals ($\delta_{ext}:S\times\Re_{\ge 0}\times X^b\to S$) and introducing a confluent transition function ($\delta_{conf}:S\times X^b \to S$) to handle conflicting situations where multiple input signals and an internal transition happen at the same time (so $\delta_{conf}$ is used instead of $\delta_{ext}$ or $\delta_{int}$). PDEVS modifies the original transition semantics to process all imminent components in parallel, reducing artificial serialization and improving performance in distributed and parallel simulation environments. PDEVS has become a common baseline for modern DE simulation frameworks due to its composability, precise time management, clear separation of concerns, and suitability for parallel execution. However, the restrictions that allow correct time advancement in DEVS or PDEVS do not reflect the high level of autonomy and loose coupling in today's asynchronous, concurrent, distributed systems, and do not support real-time requirements. We will revisit this matter in section \ref{sec:formalism} and \ref{sec:behave}.

\section{The M Toolchain} \label{sec:toolchain}
\begin{figure}[H]
    \centering
    \includegraphics[width=0.8\textwidth]{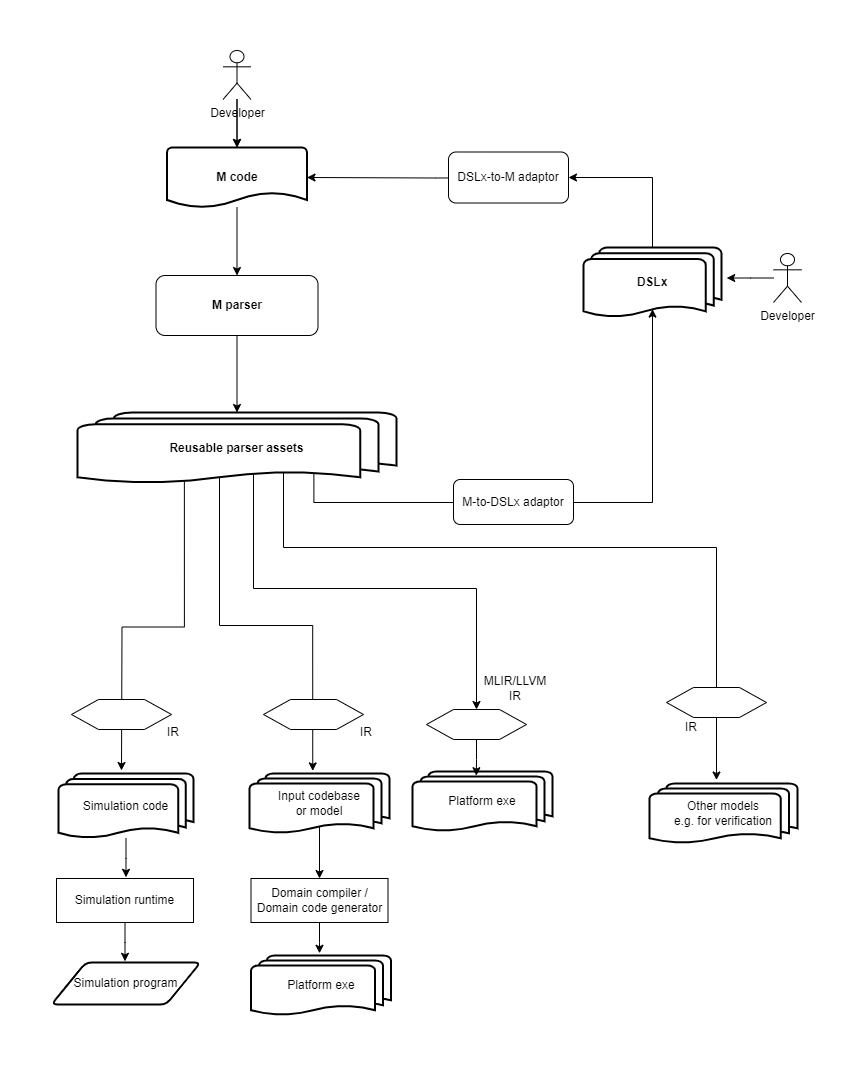}
    \caption{The M toolchain for reusable model compilation}
    \label{fig:m}
\end{figure}

In this section, we describe the M toolchain for reusable model compilation (see Fig.~\ref{fig:m}). Its workflow can be summarized as follows.

\begin{itemize}
    \item \textbf{System model as entry point:} An M system model is the starting point of the workflow. The model is parsed once to capture its structure and semantics. The resulting artifacts (e.g., parse tree, decorated abstract syntax tree (AST)) are preserved and reused by subsequent components. This avoids repeated parsing and semantic analysis, ensuring efficiency and consistency.
    
    \item \textbf{Input:} Models can enter the toolchain directly or indirectly. A \emph{direct input} is a model written in M, which functions as a standalone modeling language. An \emph{indirect input} originates from another modeling language (e.g., DSLx), which is first adapted to M. This adaptation can be performed manually (semantic analysis and translation by human interpretation) or automatically (through a dedicated DSLx-to-M model transformation).
    
    \item \textbf{Output:} The toolchain supports multiple targets, depending on the purpose: simulation code for execution on supported simulation platforms, formal verification models (e.g., Timed Rebeca, UPPAAL), codebases for domain-specific compilers or code generators, and direct executable code via the MLIR/LLVM stack \cite{mlir21,llvm}.
    
    \item \textbf{Intermediate Representations (IR):} Transformations may employ intermediate representations to abstract common structures before specialization. This aligns with the concepts of Platform-Independent Models (PIM) and Platform-Specific Models (PSM) in Model-Driven Architecture (MDA) \cite{DBLP:journals/sosym/Brown04}. For example, UML models can first be serialized to the standard UML XMI (e.g., via Eclipse UML2/Papyrus export) as a textual IR before conversion into M.
\end{itemize}

\noindent
To serve as the core enabler of this toolchain, the M language must satisfy the following requirements:

\begin{itemize}
    \item \textbf{Textual, grammar-driven design:} The language must be grounded in grammar-based syntax to leverage established language and compiler theory. It should be expressive enough for a broad range of systems, while avoiding the pitfall of attempting to cover entirely unrelated domains (e.g., finance and robotics). In consideration of research and collaboration ecosystem, we target reactive concurrent systems, cyber-physical systems, and distributed robotics.
    
    \item \textbf{Declarative nature:} The language should function primarily as a declaration or specification language, describing system structure and behavior. It should remain implementation-free, basically specifying \emph{what-to-do}, the \emph{how-to-do} part is only algorithmic description like in a pseudocode.
    
    \item \textbf{Formal semantics:} A strong formal foundation is essential to justify the processing logic and to enable precise interpretation and execution.
    
    \item \textbf{Facilitation of translation:} The language must be sufficiently simple and semantically transparent to allow accurate conversion into other modeling or implementation formalisms with semantic preservation.
\end{itemize}

We examined a wide range of existing modeling languages, including ROOM \cite{Bran94}, UML-RT \cite{umlrt98}, SysML v2 \cite{sysmlv2}, Modelica \cite{DBLP:books/daglib/0020345}, Karma \cite{karma1}, Ptolemy II \cite{ptolemy2}, Lingua Franca \cite{DBLP:journals/tecs/LohstrohMBL21}, P \cite{p}, ABS \cite{10.1007/978-3-642-25271-6_8}, ThingML \cite{thingml}, Miking \cite{miking2019}, and Timed Rebeca. None of the existing languages fully met the requirements for our toolchain, which led to the creation of M for a neater feature set, more transformability, and extensibility. It is important to note, however, that the M toolchain does not mandate the exclusive use of our M language. Other languages may serve as entry points if they better suit a given context, provided the paradigm of semantics-aware model compilation is preserved. In Section~\ref{sec:cmp}, we revisit this point by comparing M with several existing languages to highlight its distinct advantages for model compilation.
\section{The M language} \label{sec:design}
\subsection{Some Conceptual Clarification}

Computational objects can be classified according to their execution model:
\begin{itemize}
    \item Passive objects only store data and provide methods that are executed when invoked by others. They cannot initiate actions by themselves \cite{meyer1997object}.
    \item Active objects have their own thread of control, enabling them to execute autonomously and process incoming requests or messages \cite{Agha1986Actors}.
    \item Reactive objects are driven by external events; they remain idle until triggered by stimuli such as messages or environmental changes. They are central in event-driven and simulation frameworks like DEVS \cite{zeigler2000theory}.
\end{itemize}
In short: passive objects wait to be used, active objects act on their own, and reactive objects respond to events. A reference or handle to an object is a value that helps to locate the object. It can be a memory address, a process number, or web address, depending on what represents the object. A class is a template, a blueprint that defines a kind of objects. Instantiation or incarnation is the process of creating a new object out of a class and returning the created object's reference. An actor class in M could be said to be a mix of passive, active and reactive elements.

An event is something that happens at a moment in time and requires some reaction. An event is described in a data structure called event notice. In M, message and event notice are equivalent. Events, tasks, jobs, actions, reactions, responses are interchangeable terms used henceforward. The stimulating source or trigger that leads to the occurrence of an event may be external - at the arrival of some signal from another actor, or internal - due to a change in state or some time-wise condition.

State variables are data that describe a system at a time, hence state values are time-sensitive and carry temporal logics (e.g. value at a moment, value before that moment). The history of time advancement and state changes need to be recorded to know how the system state evolves over time and to detect temporal conditions. Non-state variables are supplementary information that can be hidden to the observer. The separation of state and non-state information depends what is necessary and of interest to the observer.

Time is a complicated dimension to model. The model's time or logical time is the artificial timeline in the model. The logical clock does not progress naturally like the physical clock, but depends on the applied modeling algorithm (or time model): by fixed increment in discrete time model, or by the next event in DE model, or by a numeric integrator for continuous equations. When the model is executed as a simulation, the execution time for a task is only an assumption of the real execution time. The wall-clock time measuring how fast a job is done in simulation normally is not related to the assumed execution time, unless formulated explicitly.

\subsection{Basic Design}

The design of M combines principles and best practices drawn from multiple modeling and programming communities, as well as our own experience. Its core concepts are summarized as follows:

\begin{itemize}
    \item \textbf{Actor-based architecture:} M adopts the actor model to describe distributed, concurrent, and asynchronous systems, supporting reasoning about communication and liveness. Actors can be hierarchically composed through child actors, enabling modular system construction.
    
    \item \textbf{Explicit communication entities:} A \textit{connection} is introduced as a first-class construct (currently sharing the same semantics as actors) to specify communication properties such as latency, transport mode, broadcasting, and listening behavior.
    
    \item \textbf{Timed message scheduling:} Each actor’s mailbox serves as its local future event list, containing messages tagged with timestamps that indicate their scheduled occurrence. Time progression follows a next-event semantics akin to classic DE simulation, with additional mechanisms to handle simultaneous events and preserve total order of time. All local clocks are assumed synchronized under a shared logical clock.
    
    \item \textbf{Temporal state semantics:} State variables are timestamped, forming time series that support temporal reasoning and trajectory tracking. An event scanning mechanism detects post-event state changes while preventing redundant re-triggering.
    
    \item \textbf{Language-level event control:} Constructs such as the \textit{with()} clause allow specification of message properties, while the \textit{cancel} command enables explicit manipulation of the event list.
    
    \item \textbf{Expressive type and execution system:} M provides a rich type system and imperative constructs, offering programming expressiveness comparable to general-purpose languages like C or Java.
    
    \item \textbf{Interfacing with external systems:} External functions can be declared in M with known signatures, while their implementations are provided by the execution platform.
\end{itemize}

The following example illustrates the language’s syntax and modeling approach. It models a water-tank controller in which a sensor periodically measures the tank’s water level. The controller operates a valve that opens when the level falls below a low threshold and closes when it exceeds a high threshold. The valve opening is represented as a fraction between 0 (closed) and 1 (fully open), adjustable both manually and automatically. Continuous quantities such as water level are discretized through periodic sampling.

\begin{lstlisting}[label=lst:watertank,style=CodeBlock, caption={Water tank controller}]
def TankController class[actor] {
    const inputSpeed:float = 1.0; //1 litre per second
    const maxLevel:float = 1000;
    const minLevel:float = 10;
    
    var[state] level:float=0;
    var[state] opening:float=0;
    
    do[every(1s)] readTankLevel{
            level += opening * inputSpeed;		
    }
           
    do[on(prev(level)<=maxLevel && level>maxLevel)] shutValve{	
        opening = 0;
    }
           
    do[on(prev(level)>=minLevel && level<minLevel)] openValve{
          opening = 1;
    }  
       
    do manualClose{
          opening = 0;
    }
       
    do manualOpen(userlevel:float){
          opening = userlevel;
    }
};        

main{
    var controller:actor = TankController(level:0,opening:1);
}
    \end{lstlisting}

\subsection{Formalization and Operational Semantics}\label{sec:formalism}
In this section, we formally present the structure and operational semantics of a system model in M, ensuring precise and consistent understanding and execution across different platforms.
\subsubsection{Assumptions}
Fully distributed systems have independent nodes without a global clock, processing events in nondeterministic order, which makes consistent interpretation across execution environments difficult. To bound complexity and standardize step-by-step behavior, we adopt the following assumptions: 
\begin{itemize}
    \item \textbf{Synchronized clocks}. There is no global physical clock in a fully distributed system. Each node maintains a local clock that may be unsynchronized. To avoid adding complexity, we assume that all clocks in M actors are synchronized and can therefore be represented by a single logical clock in simulation, so that time across actors can progress uniformly. It is not necessary to address clock asynchronicity and synchronization problems in this context (e.g., use a separator actor to represent a reference clock and communicate with this actor to synchronize local clocks). The only rule to observe is the \emph{happened-before} rule, or the causal ordering of events in a distributed multiprocess system, as stated by Lamport \cite{lamport1978time,vanSteenTanenbaum2017Distributed,CoulourisDollimoreKindberg2011Distributed}: the moment a message is received is always after the moment it is sent, so $t_{received} > t_{sent}$ in any case along the same timeline.
    \item \textbf{Time advancement}. As messages can be ordered chronologically on the same timeline, the logical clock can advance following the same logic of next-event time as in a DE system simulation, except for when multiple concurrent events may happen in parallel in different actors.
    \item \textbf{Tie-breaking}. In case of concurrent events, a tie-breaking mechanism is needed to determine which event should be processed first, since an M actor only handles events one by one. By default, we set the priority by \textit{information value}: the more unknown an event is, the more important it is. For example, a stop/reset signal should override all other events. Therefore, in terms of significance, the order is external events first, then conditional events, and periodic events\footnote{In rare scenarios where an actor, after filtering, still has multiple concurrent events of the same type, then a random selection is applied.}. So each time, system-wide, the M simulator knows which ones to take out depending on their timestamp, category and receiver. A user-defined tie-breaking function can be facilitated to override the default selection; however, currently it is not implemented yet because we have not found sound reasons for such policy customization.
    \item \textbf{Stochasticity}. Most random phenomena can be represented by suitable probability distributions. Comprehensive treatments are provided by Law~\cite{law2014simulation} and Banks~\cite{banks2010discrete}. Accordingly, the stochastic events in M can be modeled by sampling their future occurrence times from an appropriate distribution.
\end{itemize}
As M follows DE time progression, we initially thought of reusing a Parallel DEVS simulator like DEVS-Suite \cite{kim2009devs_suite}\footnote{\url{https://acims.asu.edu/devs-suite/}} or PythonPDEVS \cite{10.5555/2872965.2872978}. However, that was not feasible because the components in M possess greater degrees of freedom than those allowed by DEVS:
\begin{itemize}
    \item In DEVS, self-looping is not allowed -- a component in DEVS can only send messages to another component, not to itself.
    \item In DEVS, message sending (output generation) can only happen at the end of a state's lifetime, while an actor in M can send a message at any point in time.
    \item In DEVS, at the arrival of an external message, the internal state of a component has to be recomputed (using the $\delta_{ext}$ function) and re-clocked (using the time advance or $ta$ function to compute the lifetime of the new state). In M, it only depends on the response function, which may or may not change the state and its lifetime.
\end{itemize}
A simulator for M is under development, with actors implemented as web services and a central database handling message storage and selection.
\subsubsection{M Formalism and Semantics}
Some conventions:
\begin{itemize}
    \item $T=\langle a,b,[c],[d] \rangle$, then $T.a$ or $a_T$ refers to component $a$ in $T$, and $c$ and $d$ are optional components whose value is  $\varnothing$ if not given.
    \item $E=\{e\}$, then $E[\rho\to v]$ refers to a subset of $E$ with elements on which the function $\rho$ evaluates to $v$: $E[\rho \to v] = \{e|e \in E \land \rho(e) = v\}$. By convention, $E[\rho]$ is the shorthand for $E[\rho \to true]$.
    \item $min(E.a)$ refers to the smallest value of component $a$ in all elements of set $E$, $min(E.a) = min(e.a|\forall e \in E)$   
    \item $\epsilon$ is a micro-timestep (e.g.,$10^{-9}$), to help order events by their occurrences. 
    \item Given $x = \{(k,v)\in K\times V|v=x[k]\}$ with $k\in K_<$ as a collection of values with totally ordered keys, that means $x=\{(k_1,v_1),\dots,(k_n,v_n)\}\text{ such that }k_1<_Kk_2<_K\dots<_Kk_n$ then $first(x)=(k_1,v_1)$, $last(x)=(k_n,v_n)$, $prev(x,k_i)=last(x[k<_K k_i])$ is the one right before $k_i$, $insert(x,k',v') = x \cup (k',v')$ is a function that inserts a new value $v'$ at location $k'$ in $x$ with respect of the total order of keys.
\end{itemize}
A \textbf{system model} in M is a tuple $M = \langle \mathbb{T},\hat{\tau},A, \hat{Q},\teq\rangle$ where:
\begin{itemize}
    \item $\mathbb{T}=\mathbb{R}$ is the domain of logical timestamps which can be a negative and decimal number of the logical time unit. The logical timestamp is not the real time, just a ticket to order events in simulation.
    \item The global logical clock $\hat{\tau}=\{\hat{\tau}[1],\hat{\tau}[2],...\}$ is not a single value but a \textbf{sequence} of strictly increasing timestamps: $\hat{\tau}[i]\in\mathbb{T}, \hat{\tau}[i]<\hat{\tau}[i+1]\ \forall i\in\mathbb N$, starting at $\hat{\tau}[1]=-1$, then $\hat{\tau}[2]=0$, and ending at current time $now() = last(\hat{\tau})$. When the clock advances to a new timestamp $t>now()$, the new value is added to the end of the clock sequence, so $\hat{\tau}\Leftarrow t$ means $\hat{\tau}[|\hat{\tau}|+1]\gets t$.
    \item $A$ denotes the set of actors or connection objects (which also are actors). Each \textbf{actor} $a\in A$ is a tuple $a = \langle S_a,E_a,Q_a,\tau_a \rangle$, where $S_a$ represents the local state of $a$, $E_a$ represents a finite set of responses to different events as explained later in this section, and $Q_a$ is the local queue of messages of $a$.
    $M$ includes two \textbf{disjoint sets of variables}: $Var_{state}$, containing \textit{state variables} that preserve their history over time, and $Var_{nonstate}$, containing \textit{nonstate variables} that hold only the current value. 
The complete set of variables in $M$ is
\(
\mathbf{Var} = Var_{state} \cup Var_{nonstate}.
\)
Let $\mathbb{T}_{\star}=\mathbb{T} \cup \{\star\}$ where $\star$ represents any time. Then, actor states are defined by the mapping function $S:Vars\times \mathbb{T}_{\star}\to Values$ such that $S[x,t]$ represents the value of $x\in Var_{state}$ at time $t$ and $S[x,\star]$ represents the value of $x\in Var_{nonstate}$. Note that $S[x,\star]$ is undefined if $x\in Var_{state}$ and $S[x,t]$ is undefined if $x\in Var_{nonstate}$. The use of clock sequence and state values in M facilitates \textbf{event logging} and \textbf{temporal semantics}: e.g. get the previous value right before current time, $prev(S_a)=last(S_a[t<now()])$. The global state store $\hat{S}$ is a map of actor ID and timestamp to its corresponding state value at that time, so $S_a[\hat{\tau}[i]] \equiv \hat{S}[a,\hat{\tau}[i]]$. All local clocks are synchronized, advance uniformly, and are represented by one logical clock, i.e. $\forall a \in A: \tau_a \equiv \hat{\tau}$.

\item $\hat{Q}=\{m\}$ is the logical global message queue (also called the message server). Since we have synchronized clocks, local message queues of actors can be projected on $\hat{Q}$ and aligned on the same timeline $\hat{\tau}$ by increasing order of $m.t_{assume}$, $\forall a:Q_a\equiv\hat{Q}[a]$. Any change in a local queue due to an event, local state, time and its current content is equally reflected in the corresponding slot in the global queue. $\hat{Q}[a]$ is shorthand form of $\hat{Q}[receiver=a]$.
    \item A \textbf{message} in $M$ is a tuple $m = \langle \text{sender}, \text{receiver}, \text{ev}, t_{\text{assume}}, [\theta], [t_{\text{send}}], [t_{\text{actual}}] \rangle$, where:
\begin{itemize}
    \item $\text{sender} \in A$ is the actor initiating the message,
    \item $\text{receiver} \in A$ is the actor receiving the message,
    \item $\text{ev}$ denotes the event type,
    \item $\theta$ represents optional message parameters,
    \item $t_{\text{send}}$ is the timestamp when the message is sent (optional),
    \item $t_{\text{assume}} = t_{\text{send}} + \Delta t$ is the assumed delivery or event occurrence time, with $\Delta t > 0$ representing transport latency or scheduled delay, as governed by Lamport's logical clock,
    \item $t_{\text{actual}}$ is a reserved field for the actual delivery time, currently set to $t_{\text{actual}} = t_{\text{assume}}$, pending future semantic rules for handling discrepancies.
\end{itemize}
  %  A \textbf{message} in M has the form $m=\langle sender, receiver, ev, t_{assume}, [\theta], [t_{send}],[t_{actual}] \rangle$ where $sender$ is the sending actor, $receiver$ is the recipient actor, $ev$ is event type, $\theta$ is the parameters of the message, $t_{send}$ is the time it is sent, $t_{assume}$ is the time it is assumed to arrive or happen, $t_{assume} = t_{send} + \Delta t$ where $\Delta t>0$, as stated by the Lamport clock. The $\Delta t$ represents the transport time, any delay, or future happening time. $t_{actual}$ is a reserved field for future use in case there is a semantics rule to handle discrepancy between the actual and the assumed happening times, for now $t_{actual} \equiv t_{assume}$.
    \item An \textbf{event response} in an actor $a$ is a tuple: $ev_a=\langle name,trigger,\sigma,T,\delta \rangle$ where $name$ defines the event type to match the received message, $trigger$ is the type of stimuli ($external, periodic, conditional$), $\sigma:(S_a,\hat{\tau}_a)\to true|false$ is the firing condition function, $T$ is the repeating interval (=0 if not periodic), and $\delta:(S_a,Q_a,\hat{\tau})\to (S'_a,Q'_a,\hat{\tau})$ is the transition function that can modify the local state and event queue.
    
    \item \textbf{Tie-breaking function} $\teq: M_{min}^b=\{m| m.t_{assume}=min(\hat{Q}.t_{assume}), m \in \hat{Q}\} \to \{m~|~|\{m \in M_{min}^b|m.receiver=r\}|\leq 1,\forall r \in A\}$ filters a bag of messages having the same timestamp to produce a list with at most one message per actor.
\end{itemize}

When actor $a$ \textbf{sends} a message to actor $b$ (which can be $a$ itself) with content $\theta$ and planned or anticipated to arrive or happen after $\Delta t$, a new message is created and put in the queue:

$m \gets \langle a, b, ev, now() + \Delta t, \theta \rangle$

$\hat{Q}' \gets \hat{Q} \cup m$ (equally, $Q'_b \gets Q_b \cup m$)

When actor $a$ \textbf{cancels} a scheduled event $m$: $\hat{Q}' \gets \hat{Q} \backslash m$ (that means existing event $m$ is removed from the queue or marked to be ignored in case removal is not allowed, e.g. in the real system)

At \textbf{start}: $\hat{\tau} = \{-1\}$, a fictional message $initialize$ is placed on the queue for all actors to help arrange the system to a required starting state, and first instances of periodic events are scheduled to come:
\\$\forall a\in A: \hat{Q} \gets \hat{Q} \cup \langle a,a,"initialize",0 \rangle$
\\$\forall a\in A, ev_a\in a.E[trigger="periodic"]: \hat{Q} \gets \hat{Q} \cup \langle a,a,ev_a.name,0 + ev_a.T \rangle$

A \textbf{loop} $while(|\hat{Q}|>0)~loop\{steps\}$ iterates through the time, in each iteration:
\begin{enumerate}
    \item Select messages with the smallest timestamp out of the message server, use the tie-breaking function to filter only one for each actor:\\ $t_{next} \gets min(\hat{Q}.t_{assume}), E_{next} \gets \teq(\hat{Q}[t_{assume}=t_{next}]), \hat{Q}\gets \hat{Q} \backslash E_{next}$
    \item \textbf{Advance the logical clock} to the next event time iff it is a future time:\\ $\text{if}\ t_{next}>now()\ \text{then}\ \hat{\tau}\Leftarrow t_{next}\ \text{else}\ \hat{\tau}\Leftarrow now()+\epsilon$
    (the clock always advances at least by $\epsilon$ to respect the total order of time)
    \item Execute in parallel reactions in  involved actors:\\ $do(m.receiver,a.E[name=m.ev]), \forall m \in E_{next}$\\
    So the state and message queue may be updated: $\langle \hat{S},\hat{Q},\hat{\tau} \rangle \to \langle \hat{S}', \hat{Q}', \hat{\tau}\rangle$. Details are described below. 
\end{enumerate}

Upon \textbf{response} to an event $ev_a$ in an actor $a$: $do(a,ev_a)$, the following happens:
\begin{enumerate}
    \item Update local state and event list: $\langle S_a,Q_a,\hat{\tau} \rangle \to \langle S'_a,Q'_a,\hat{\tau} \rangle = ev_a.\delta(S_a, Q_a, \hat{\tau}) \equiv \langle \hat{S},\hat{Q},\hat{\tau} \rangle \to \langle \hat{S}',\hat{Q}',\hat{\tau} \rangle$, which means that during an event reaction, a component may update its local state and event list (or leave them unchanged), with corresponding updates to the global system state and message server. The simulation clock remains fixed during the reaction and is advanced only to the event’s timestamp immediately before processing—this is the defining principle of DE simulation.

    \item Scan for changes:\\ $\forall ev \in E_a[trigger="condition" \land \sigma(S'_a,\hat{\tau}) \to true] :  a.send(a,ev.name,now()+\epsilon)$
    \item Schedule next iteration if $ev_a$ is a periodic event:\\ $a.send(a,ev_a.name,now() + ev_a.T) \iff ev_a.trigger="periodic"$
    
\end{enumerate}

\textbf{Terminal condition}: the system stops evolving when $|\hat{Q}|=0$, i.e. there are no more incoming messages for any actor. So when a $terminate()$ function is called to halt execution, the global queue is emptied, thus all local queues are emptied too: \\$\hat{Q} \gets \emptyset \equiv \forall a\in A: Q_a \gets \emptyset$.

\subsection{Syntax Design}
Listing \ref{lst:ebnf} presents the syntax grammar of M in EBNF format\footnote{We use ANTLR-like style with : for rule definition, * for zero or more, + for at least one, ? for at most one, $\vert$ for alternatives, \underline{under} for terminal strings, CAP for lexical rules. This is not a working grammar for the parser, but just to help readers understand the structure and syntax of a program in M.}.
%put EBNF here so that it is on one page only
\begin{multicols}{2}   
\begin{lstlisting}[label=lst:ebnf,style=EBNF, caption={EBNF grammar of M}]   
program: annotation* includeStm* constDecl* functionDecl* typeDecl* mainBlock

annotation: '@' ID properties? ('{=' .* '=}')?
    properties: '[' property (',' property)* ']'
    property: ID (':' valExpr)?

includeStm: 'include' stringVal ';'
constDecl: 'const' ID ':' type '=' expr ';'    
varDecl: 'var' properties? ID ':' type ('=' expr)? ';'    
typeDecl: 'def' ID  singleType|collectionType|tupleType|recordType|classType ';'
type: 'any'|singleType|collectionType
singleType: numeric|'bool'|'char'|'string'|'timespan'|ID
collectionType: 'array'('{'type (',' INT)?'}')? 
    | 'list'('{'type'}')?
    | 'set'('{'type'}')? 
    | 'map'('{'type','type'}')?
tupleType: 'tuple' '{'type (','type)*'}'
recordType: 'record' '{'(ID ':' type ';')+'}'
classType: 'class' classProperties? ('extends' idList)? '{' classMember* '}'
classProperties: '[' 'object'|'actor'|'connection' ']'
classMember: varDecl|constDecl|functionDecl|doDecl

doDecl:'do' doProperties? ID ( '(' formalArgs? ')' )? block
    doProperties: '[' On|Every|Receive ']'
    On: 'on' '('expr')'  (',' property)*
    Every: 'every' '('timespanVal|ID')'  (',' property)*
    Receive: properties? //'receive' is skipped

mainBlock: 'main' block
block: '{' statement* '}'
statement: varDecl
    |expr';'
    |'return' expr?';'
    |'if' '('expr')' block ('else' block)?
    |'cases' '('expr')' '{' ('case' expr ':' statement* ';')+ ('otherwise' ':' statement* ';')? '}'
    |'while' '('expr')' block
    |'do' block 'while' '('expr')'';'
    |'foreach' '(' 'var' idList 'in' 'values'|'keys'|'pairs''('expr')'')' block
    |'break'';'
    |tellStm
    |cancelStm

tellStm: ID '!' ID ('(' exprList ')')? withClause? ';'
    withClause: 'with' '(' idExprList ')'

cancelStm: 'cancel' '*'|idList ';'

expr: assignExpr|ifExpr|caseExpr|binaryExpr|prefixExp|postfixExpr|valExpr|'('expr')'

valExpr: standardLiterals|timespanVal|collectionVal|tupleVal

timespanVal: timeUnit+
    timeUnit: DECIMAL ('ms'|'us'|'ns'|'h'|'m'|'s'|'d')

collectionVal: ('array'|'list'|'set''('exprList?')') 
              |('map''('idExprList?')')

tupleVal: 'tuple''('expList')'

postfixExpr: 'self'|ID trailerExpr*
    trailerExpr: ('.'ID) | ('['expr']') | ('('argList?')')
    argList: exprList | idExprList 

functionDecl: extFunc | intFunc   
    extFunc: 'external' 'function' ID '('formalArgsT?')' (':' typeT)? ;    
    intFunc: 'function' ID '('formalArgs?')' (':' type)? block
    
idList: ID (',' ID)*
exprList: expr (',' expr)*
idExprList: ID ':' expr (',' ID ':' expr)*
formalArgs: ID ':' type (',' ID ':' type)*  
typeT: type|'#' ID;
formalArgsT: ID ':' typeT (',' ID ':' typeT)*  
\end{lstlisting}
\end{multicols}
M follows the principle of \textit{"semantics first, syntax later"}. Its syntax reflects the M formalism while bearing the look-and-feel of popular programming languages like Java, C, PHP. The design in general adheres to following principles:
\begin{itemize}
    \item Single way of doing things: redundant constructs are generally avoided, e.g. a $for$ loop is not provided as it can always be equally done by a $while$ loop.
    \item Balance between human friendliness and parser friendliness: M is designed to be readable, comprehensible, editable for humans. At the same time, the syntax must care for avoiding ambiguity and easing pattern recognition.
    \item Distinction and consistency: M syntax facilitates easy pattern recognition through dedicated starters and bracketed modifiers (e.g., $var$[$state$], $class$[$actor$], $do$[$every$(...)]).    
    \item Extensibility: we start with a skeletal version but prepare for future extensions (see more in \ref{sec:extend}).
\end{itemize}

Verbally explained, a program in M is an executable system model which contains:
\begin{itemize}
    \item Header annotations, which may contain extra information to tell the compiler how to execute the model or how to generate the resulting code.
    \item Library files included through the $include$ command.
    \item Configuration parameters as shared constants
    \item Definitions of actor classes and other helpful data structures.
    \item A $main$ function where the actors are instantiated. The activation and initialization of actors are done hiddenly by the execution framework.    
\end{itemize}

A class in M, defined with \texttt{def\ A\ class[actor|connection]}, contains:
\begin{itemize}
    \item Local variables which can be state or non-state.
    \item Local constants for convenient local references.
    \item Local functions: equivalent to ordinary private methods in a passive object class.
    \item Event types and handlers are specified with the $do$ construct. The type and parameters of an event are conveniently specified through the name and formal arguments of a $do$. The triggering type, firing condition and other properties are embedded in the $[properties]$ part of a $do$.
\end{itemize}
M provides syntactic sugars for conveniently accessing time and state variables, like:
\begin{itemize}
    \item $now()$ returns the current time = $last(\hat{\tau})$.
    \item Current value of a state variable $s$ is represented by $s$, not $current(s)$, though underneath the compiler will return the last recorded value of $s$, as in the semantics. And the assignment $s = s_1$ is actually implemented as $s[now()] \gets s_1$.    
    \item $prev(s)$ returns the value of state variable $s$ right before $now()$. When the clock advances, $now() \to now() + \Delta t$, $prev(s) = last(s[t<now()])$ returns the current value of $s$, which prevents the old conditions to reoccur, avoiding refiring happened events.
\end{itemize}

\subsection{Modeling of System Architectures}
In a broader sense, a system is a collection of entities which can be humans, machines or software computational units that interact with each other over time towards the accomplishment of some logical end (see \cite[ch.~1, p.~3]{law2014simulation}, \cite[ch.~3, p.~91]{banks2010discrete}). Kazman et al. \cite[ch.~I]{DBLP:journals/sigsoft/Herzog15} defines "the software architecture of a system is the set of structures needed to reason about the system, which comprise software elements, relations among them, and properties of both." So to model system and software architectures, entities and relationships are captured in M through actors, connection channels and messages. Example \ref{lst:talker} explains the concepts.

\begin{lstlisting}[label=lst:talker,style=CodeBlock, caption={System architecture modeling in M}]
include "mcore.m";
//define Talker as a class of actor
def Talker class[actor] {
    var name:string;
    var ch:connection; //handle of the broadcast channel

    //"initialize" is the first message in the queue
    do initialize{
        ch!register(self); //tell ch to register this talker
        self!talk() with(after:3s);
    }
    //every 3s it will trigger talk()
    do[every(3s)] talk(){
        ch!send("Do you hear me?");
    }

    do hear(string msg){
        println("I, ${name}, heard ${message.sender} said ${msg}!");
    }
};

//define BroadcastChannel as a connection class
def BroadcastChannel class[connection]{
    var name:string;
    var talkers:map{actor,bool}=map();
    
    do register(whom:actor) {
        talkers[whom] = true;
    }

    do send(string msg){
        foreach(var who:actor, registered:bool in entries(talkers)){
            if (registered){
                who!hear(msg) with(sender:message.sender);
            }
        }
    }
};

main {
    var ch1:connection = BroadcastChannel(name:"channel1");
    var talkers: array{actor}; //a dynamic array of Talker actors
    //create talkers
    var i:int = 1;
    while(i<=10) loop {
        talkers.add(Talker(name:"talker #${i}",ch:ch1));
        i++;
    }

    //initialize and run are done implicitly by the executor        
}
\end{lstlisting}

\subsection{Modeling of System Behaviours}\label{sec:behave}
In a word, system behaviours are defined by \textit{when-to-do-what}. M provides the \textit{do} construct with different triggering conditions to express this view:

\begin{lstlisting}[style=CodeBlock, caption={The \emph{do} construct in M}]
do[every(timespan)] periodicEventName { ... }

do[on(condition)] conditionalEventName { ... }

// when receive an external message named M1 the [receive] specifier is omitted
do M1(args){ ... }
\end{lstlisting}

The syntax of the $do$ construct helps conveniently define corresponding message structure. An event notice or message in M is encapsulated in an envelop which contains:
\begin{itemize}
    \item Event type or task name, which is matched with the name of a corresponding $do$,
    \item Content (payload) fields, which are fed as actual values to formal arguments of the $do$,
    \item Other timing, routing and behavioral properties are given in an optional $with()$ clause when the message is sent with the $tell$ command.
\end{itemize}
The \textit{tell}\footnote{We name it $tell$, not $send$, because $tell$ is more semantically meaningful, as in "tell somebody to do something"} operator (!) allows an actor to send an instant or future message to itself or others with content parameters in function arguments and extra parameters in a $with$ clause. The \textit{cancel} command allows purging/ignoring all or several particular event types which are present in the queue. M has a \textit{timespan} type that allows to specify time measurement in different units, e.g. $1.1d2.2h3m4s5ms6us7ns$ means 1.1 days, 2.2 hours, 3 minutes, 4 seconds, 5 mili-seconds, 6 micro-seconds, 7 nano-seconds. All $timespan$ values will be converted to a decimal of a common unit defined by a constant named \texttt{SIM\_TIME\_UNIT} (see \ref{sec:timespan}).

The code in Example \ref{lst:cfm} models a coffee machine to demonstrate event scheduling and cancellation semantics: a simple coffee making machine has one button to start or stop pouring coffee into a cup of maximum 150ml. It does not pour all in once but 5ml every 100ms (so 150ml in 3s), so that the user can stop midway if they want to have less. This example demonstrates step-wise discretization of a continuous activity and cancelability of an on-going action.

\begin{lstlisting}[label=lst:cfm,style=CodeBlock, caption={The coffee machine model}]
include "mcore.m";

def Drippper class[actor]{
    const dripSpeed:real=5; //5ml
    const dripPeriod:timespan=100ms;//every 100ms
    var[state] drippedAmount:int=0;
    var controller:actor = null;//reference to the controller
    
    do drip(amountMax:int){
        drippedAmount += dripSpeed;
        if (drippedAmount >= amountMax) { self!stop; }
        else { self!drip(amountMax) with(after:dripPeriod); }
    }
    
    do stop {
        cancel drip; //cancel the scheduled dripping action
        drippedAmount = 0;
        controller!done; //tell the controller to do "done"
    }
};

def Controller class[actor]{
    var[state] busy:bool=false;   
    var dripper:actor; //refer to the dripper unit
    
    do startstop {
        if (!busy) { busy = true; dripper!drip(150); } 
        else { dripper!stop; }
    }
   
    do done { 
        busy = false; 
    }

    do initialize{
        //assume a button press after 5s
        self!startstop with(after:5s);
    }
};

main {
   var theController:actor = Controller(dripper:theDripper);
   var theDripper:actor = Dripper(controller:theController);      
}        
\end{lstlisting}

We accept the below assumptions and limitations, which on the other hand is the working mechanism of the classic DE time model:
\begin{itemize}
    \item The logical clock only advances by event timestamps, not by the execution time of any response. To model execution time, a fictional event has to be created to mark the completion of a task.
    \item Similar for deadline missing (when a message arrives later or an action takes longer time than expected), we have to mark it as another event because in the DE time model, events are always executed at the presumed time.
\end{itemize}
The lack of built-in semantics for modeling execution time, deadline imposition implies complication in modeling real-time requirements with the classic DE time model. We will investigate this matter in a separate research because it may result in modifying the DE time model or applying a new time model rather than DE. For this stage, we deliberately reduce the complexity of the model to move faster with a prototype.

\subsection{Modularity}
Several mechanisms are designed to organize the code and construct the components in M in a more modular, hierarchical, compositional way:
\begin{enumerate}
    \item File splitting and inclusion: common definitions of constants, types and functions can be split to a separate file to be included in different models. This file inclusion works similar to the $include$ command in C or PHP.
    \item Library wrapping and symbol importing: We also reserve a $library$  entity and a $use$ command to wrap and import definitions in a library. The reason for this construct is that in distributed deployment, there are no shared codes, common definitions have to be wrapped up and deployed to each node.
    \begin{lstlisting}[style=CodeBlock]    
    //in "common.m" file
    library common{
        def real double;
        const pi:real=3.14;
        external function power(x:real,n:real):real;        
    }

    //in main file
    include "common.m";
    use common.*;    
    \end{lstlisting}
    \item Object-oriented composition: A simple object-oriented style of class inheritance and polymorphism can be implemented to help build actor classes in a more reusable, modular way.
\end{enumerate}
Currently only (1) is provided, other mechanisms are planned for future releases.
\subsection{Extensibility}\label{sec:extend}
External functions, properties and annotations are effective mechanisms to extend the semantics of M. An \textit{external function} has signature declared in M while realization is provided by the execution platform. The compiler parses an external function's signature to know its applicable types, formal arguments and return type. Example:
\begin{lstlisting}[style=CodeBlock]
    external function prev(s:#T):#T;    
    external function pow(x:real,y:real):real;
    external function toString(a:#T):string;
    external function push_back(a:#T,e:#E):#T;    
    var t:real = pow(10,5);//10^5
    var a:array{int} = array(1,2,3);
    a = reverse(a); //(3,2,1)
    push_back(a,4); //(3,2,1,4)    
    toString(a);
\end{lstlisting}
This is only an initial design decision, we will revise the constructs for pattern definition and matching in the next revision.

Extra properties can be added to the property list within [...] of different lexical types. E.g.: variables may have $discretize$ property, so $var[state,discretize:"real2int"]\ x:real$, or a new type of $do$ can be added, e.g. when missing the expected deadline of a message: $do[miss]~M1~ \{...\}$

Annotations can have properties and a free text body $@annotate[properties]\{= .* =\}$, and can be added to different parts to provide extra directive information to the compiler.

Extra parameters of messages can be added to the $with$ clause, e.g. a $reply$ field to tell the recipient actor to send back the result to: $other!doSmth(1,2,3)~with(reply:"onReply");$ 

\subsection{Towards Hybrid System Modeling}
A distinctive feature of the M semantics is its treatment of clocks and state variables as \emph{time-indexed series}, which provides a natural basis for supporting \emph{numerical integration methods} \cite[ch.~11, 17]{Fishwick2007} that approximate continuous-time dynamics within the DE semantics of M. In particular, many physical systems exhibit continuous evolution, typically described by differential equations of the form
\[
\frac{dx}{dt} = f(t, x), \quad x(t_0) = x_0.
\]
Within M, such dynamics could be represented by discrete samples of the trajectory \(x(t)\), obtained by applying numerical solvers such as Euler’s method, Runge–Kutta methods, or adaptive step solvers. These solvers iteratively compute approximate values of \(x\) at successive time milestones of the clock \(\hat{\tau}\), thus producing a discrete representation of the continuous evolution. The logged state series in M directly supports storing and reasoning about these trajectories. This extension would effectively elevate M into the realm of \emph{hybrid system modeling}, 
where both DE behaviour and continuous dynamics (physical processes, control systems) coexist under a unified semantic 
framework. Such an integration offers several benefits:

\begin{itemize}
  \item \textbf{Unification}: hybrid behaviour can be expressed in the same language, 
        avoiding the need to couple external simulation tools.
  \item \textbf{Semantic clarity}: continuous dynamics are discretized in a manner consistent 
        with the underlying discrete-event semantics of M.
  \item \textbf{Multi-target compilation}: solvers for continuous dynamics can be selected 
        at code generation time, adapting to the requirements of simulation, verification, 
        or deployment.
\end{itemize}
However, several challenges arise. Step-size management must be reconciled with the event scheduling semantics, interpolation between sampled values must be well-defined, and approximation errors must be considered when reasoning about correctness. We therefore regard this direction as a promising future extension of M, opening the possibility of supporting cyber–physical and hybrid models in a principled way.
\section{Type System of M}\label{sec:types}
%\subsection{Introduction and Motivation}
A type system is a categorization scheme (taxonomy) for information holders. The design of a type system is central to any programming or modeling language. Types define the kinds of values that can be expressed, the operations permitted on them, and the guarantees that the compiler or analysis tools can provide. In the M language, the type system has been designed not only for traditional concerns such as safety and efficiency, but also to support \textit{modeling and analysis} in concurrent, distributed, and time-aware systems. Unlike general-purpose programming languages, M is aimed at \textit{model compilation} and \textit{formal verification} contexts, where precise type distinctions directly affect correctness. A model is not simply an \textit{executable artifact}; it is also a \textit{mathematical object} that must be analyzable, simulatable, and transformable into other formalisms. For this reason, M’s type system balances among various competing requirements:
\begin{enumerate}
\item \textbf{Expressiveness for modeling domains.}
M must allow the definition of domain-specific types that capture entities such as physical quantities, communication channels, actors, and state machines. The type system provides structured types ($records$, $tuples$, $classes$) and extensible class kinds (e.g., $class[actor]$, $class[connection]$) to support modeling abstractions directly.

\item \textbf{Safety and analyzability.}
The type system enforces static checks to prevent inconsistencies: invalid assignments, mismatched collection types, unresolved type inference, or illegal conversions. This ensures that models remain analyzable and that transformations (e.g., into automata or state-transition systems) preserve semantics.

\item \textbf{Clarity and reuse.}
Aliases and user-defined types allow modelers to capture domain-specific semantics while reusing underlying primitive or structured types. For example, defining \texttt{def Meter double} gives semantic clarity in physical models while still mapping to efficient scalar representations.

\item \textbf{Support for automatic inference.}
Syntax of declarations and literal values in M is designed to be self-describing, enabling the compiler to infer each symbol's type (and element type, collection size for collections) where possible. At the same time, unresolved or ambiguous typing is flagged as a semantic error, ensuring that models are never underspecified.

\item \textbf{Compatibility with analysis backends.}
Since M models may be translated into formal verification frameworks (such as timed automata, process algebras, or state-transition systems), the type system provides a bridge between high-level expressiveness and low-level formal rigor.
\end{enumerate}

In short, the type system of M is motivated by the need to be expressive enough for modeling complex systems, precise enough for executability and formal verification, efficient enough for compilation and execution while staying friendly enough to developers.
\subsection{Type Hierarchy of M}
\begin{figure}[H]
    \centering
    \includegraphics[width=1\linewidth]{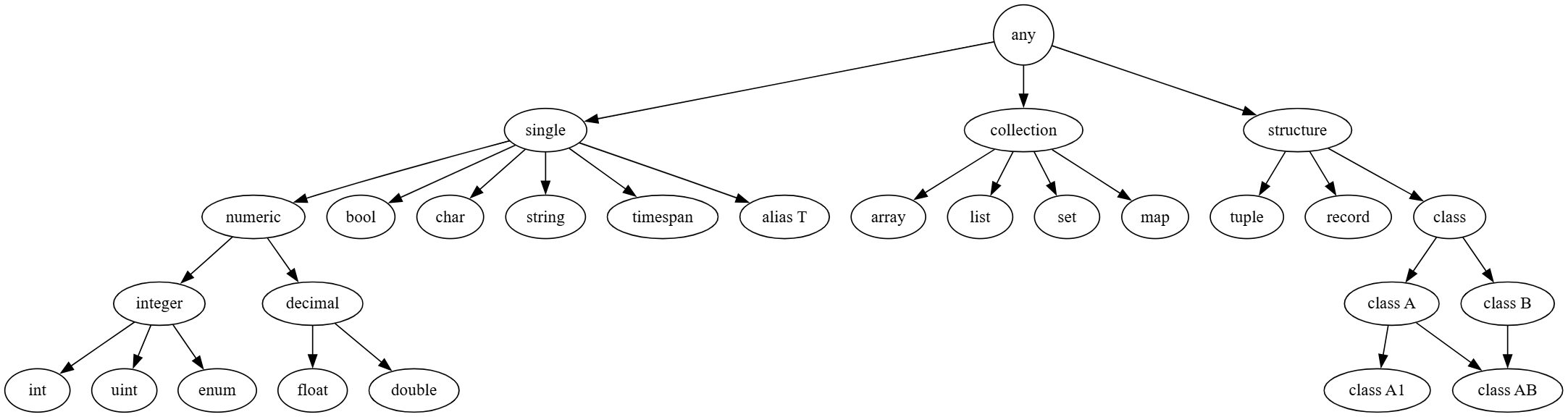}
    \caption{Taxonomy of M type system}
    \label{fig:types}
\end{figure}

M provides primitive scalar types, collection types, user-defined structured types (tuples, records, classes), and alias (see Fig.\ref{fig:types}). These types are equivalently supported natively or through workarounds in implementation languages like C++ or Java.
Type definitions follow this syntax pattern which reads as "define T as ...":  \[\text{def}~T~s\text{impletype}|\text{complextype}\{\text{spec}\};\] Specification of a complex type is wrapped in "$\{spec\}$" while instantiation is "$\text{typeID}(args)$", similar to the style used in object-oriented programming. Variables and constants are declared with explicit static type: {$var|const\ ID:type = value;$}

    \begin{lstlisting}[style=CodeBlock]
def A array{int,3}; //define A as array of int with size of 3
var a:A = array(1,2,3);

def Person record{name:string;age:int;};
var b:Person = Person(name:"Johan",age:30);

def Machine class[actor]{
    var name:string;
    var[state] busy:bool;

    do initialize{ busy=false; }
    do start{busy=true;}
    do stop{busy=false;}
}
var c:actor = Machine(name:"m1");
    \end{lstlisting}

Implicit type inference is doable as types of literal values in M are recognizable from their lexical values. However, we reserve it as a future improvement. With types that can be determined at compile time through explicit declaration or implicit inference, a model in M and its derived programs can be safe and sound in terms of type, therefore be more semantically correct and executable. 
\subsubsection{Late Binding}
M has a keyword $any$ to represent a type placeholder which will be later resolved by the compiler through semantics analysis. If a symbol's type remains as $any$ after semantics analysis is done, i.e. its type cannot be deterministically decided at compile time, the compiler will output a semantic error. Types in M have to be resolved during compile time because the compiler will need to convert M to other languages that normally require strong typing for correct execution.

\subsubsection{Single Types and Collection Types}
A single type can be a primitive scalar type or an alias referring to a user-defined type. Primitive scalar types in M are: $int/uint(8|16|32|64)$\footnote{M targets multiple platforms which can be 8-64 bits, so no default word length}, $enum$, $float$, $double$, $bool$, $char$, $string$, $timespan$. Except for $timespan$, others are similar to C or Java, we therefore only show some examples. A collection type is a container of items of a single type. Collection types can be $array$, $list$, $set$, key-value $map$. The type constructors and literal value formats are elaborated as below:

\textbf{Scalar types}:
\begin{lstlisting}[style=CodeBlock]
    def int int64; //64-bit platform
    def uint uint64;
    def real double;    
    def modes enum{OFF=0,ON};
    
    var i:int=1;
    var x:real=1.23e+4;
    var b:bool=true;    
    var m:modes=modes.ON;
    var ch:char='a';
    var s:string="abc";
    var t:timespan=1.1d2.2h3m4s5ms6us7ns;     
\end{lstlisting}%

\textbf{Array}: an array in M is an indexed sequence of elements of the same type. If array size is defined, it is a fixed-size array, if not it is a dynamically allocated array.
\begin{lstlisting}[style=CodeBlock]
    def A array{int,3}; //an array of 3 integers
    var a:A = array(1,2,3);
    var a2:A = [1,2,3]; //shorthand form    

    def B array{string}; //an array of strings with dynamic size
    var b:B = array("a","b","c);
    var b2:B = ["d","e"];    
\end{lstlisting}

\textbf{List}: a list in M is an double-linked list which is a more dynamic data structure than an array for holding an ordered collection that keeps changing frequently.
\begin{lstlisting}[style=CodeBlock]
    def L list{int};
    var l:L = list(1,2,3);
\end{lstlisting}

\textbf{Set}: a set in M is a set of unique elements of a type.
\begin{lstlisting}[style=CodeBlock]
    def S set{int};
    var s:S = set(1,2,3);
\end{lstlisting}

\textbf{Map}: a map in M is a key-value hash table that maps a key to a value.
\begin{lstlisting}[style=CodeBlock]
    def M map{string,int};
    var m:M = map("a":1,"b":2,"c":3);    
    var m2:M = {"d":4,"e":5}; //shorthand form
\end{lstlisting}

Collection types can be declared without a specification, instead the compiler will try to infer during compile time, if remaining undecidable, a type error is produced. E.g.: 
\begin{lstlisting}[style=CodeBlock]
    var a:array; //equal to array{any}
    a = array(1.1,2,3); //then type of a is array{double}
\end{lstlisting}

Common methods for handling collections like \textit{keys}(), \textit{values}(), \textit{pairs}(), \textit{add}(), \textit{removeAt}() ... are provided as external functions. A \textit{foreach} loop is provided to iterate through a collection by its key, value or key-value pair sets.
\begin{lstlisting}[style=CodeBlock]
    var a:array{int} = array(1,2,3);
    foreach(var v in values(a)){
    ...
    }
    var m:map{string,int} = map("a":1,"b":2,"c":3);
    foreach(var k,v in pairs(m)){
    ...
    }
\end{lstlisting}
\subsubsection{Structured Types}
Structured types in M are $tuple$, $record$, and $class$. A tuple is a fixed sequence of items of predefined types. A record is a tuple with field names. A class in M can have extra properties indicating its kind or formalism, e.g. $class[object],class[actor], class[connection]$. Similar to Modelica style, M does not need a $new$ keyword to instantiate a class or record. 

    \begin{lstlisting}[style=CodeBlock, caption={Declaration and instantiation of structured types}]
    def T tuple{bool,int,string};
    var t:T = tuple(true,1,"test");    

    def Person record{name:string;age:int;};
    var p:Person = Person(name:"Johan",age:30);

    def A class[actor]{
        var[state] level:double;              
    }
    var a:actor = A(level:0.0);
    \end{lstlisting}

References to class instantiations can be $object$, $actor$ or $connection$. An $object$ is a memory address of a passive object which resides in memory, waiting to be invoked, like objects in C++ or Java (no state variables or event reactions). An $actor$ reference can be a memory address, process number or web address, depending on realization form of the execution platform. Actors just know each other by their references to send messages to, and there is no need to cast from one actor class to another. $connection$ normally only exists in simulation and is skipped in deployment because it is provided by the underlying communication protocol and library. The syntax grammar of M is designed to help easily recognize types from declarations and lexical values.

\subsection{Type Environment}
In the formal semantics of M we use a type environment, denoted by 
$\Gamma$, to capture the typing assumptions available in a given context.
The environment is treated abstractly as a mapping between program identifiers 
and their associated type information. Its precise implementation 
(e.g.\ hash tables, scope objects, symbol tables) is left unspecified. 
Conceptually, $\Gamma$ provides the following:

\begin{itemize}
  \item \textbf{Variable bindings}. $\Gamma(x) = T$ means variable $x$ has type $T$.
  
  \item \textbf{Function and method signatures}. $\Gamma(f) = (T_1, \ldots, T_n) \to T$  
      means $f$ takes arguments of type $T_1 \ldots T_n$ and returns value of type $T$.

  \item \textbf{Type aliases and user-defined types}. $\Gamma(\mathsf{Alias}) = T$ means alias $\mathsf{Alias}$ refers to underlying type $T$.

  \item \textbf{Structured types}. 
    For tuples, records, and classes, $\Gamma$ records the type of each named 
    field or indexed position.

  \item \textbf{Scopes}. 
    The environment may be extended or restricted to model block scopes, 
    class scopes, or library/module scopes. For example: $\Gamma, x:T$ denotes an environment where $x$ is newly bound to $T$.
\end{itemize}
Typing judgments are always relative to $\Gamma$, $\Gamma \vdash e : T$ states that, under the assumptions in environment $\Gamma$, expression 
$e$ has type $T$.
\subsection{Type Operations}
The type system of \textsc{M} defines a set of relations and operators on types
that are used in typing rules, coercion, and inference. 
We summarize the main operators in Table~\ref{tab:type-ops}.

\begin{table}[h]
\centering
\resizebox{1.0\textwidth}{!}{% <------ Don't forget this %
\begin{tabular}{ll}
\hline
\textbf{Operator} & \textbf{Meaning} \\
\hline
$T_1 = T_2$ & Type equality (nominal or structural depending on $T$). \\
$T_1 \neq T_2$ & Inequality of types. \\
$T_1 | T_2$ & Type $T_1$ or type $T_2$. So $any|T=T$. \\
$S \;\sqsubseteq\; T$ & Assignment compatibility: a value of type $S$ can be assigned to a variable of type $T$. \\
$S \;\hookrightarrow\; T$ & Implicit coercion: safe widening conversion from $S$ to $T$. \\
$S \;\Rightarrow\; T$ & Explicit conversion (cast): may be unsafe or lossy. \\
$T_1 \sqcup T_2$ & Type unification: least common supertype of $T_1$ and $T_2$ (if defined). \\
$S \;\leadsto\; T$ & Type S can be converted to type T either implicitly or explicitly. \\
$\Theta(e:T)$ & \makecell{A context $\Theta$ (e.g. user-type definition, assignment, function call, operator argument)\\ that requires element $e$ to have type T.}
\\
$\Theta[e]$ & Value of $e$ in context $\Theta$.
\\
$\llbracket e \rrbracket, \mathrm{val}(e)$ & Actual (run-time) value of $e$.
\\
$\mathrm{lval}(e)$ & Lexical value of $e$.
\\
$\Gamma(e)$ or $\mathrm{type}(e)$ & Current type of $e$ resolved in the type environment $\Gamma$.
\\
$\mathrm{elem}(T)$ & Element type of a collection $T$ (array, list, set). \\
$\mathrm{key}(M)$, $\mathrm{value}(M)$ & Key and value types of a map $M$. \\
$\mathrm{proj}(T,i)$ & Type of the $i$-th element of a tuple $T$. \\
$\mathrm{field}(T,f)$ & Type of field $f$ of a record or class $T$. \\
$\mathrm{isPrimitive}(T)$ & Predicate: $T$ is a primitive scalar type. \\
$\mathrm{isNumeric}(T)$ & Predicate: $T$ is a numeric scalar type. \\
$\mathrm{isCollection}(T)$ & Predicate: $T$ is a collection type. \\
$\mathrm{isStructured}(T)$ & Predicate: $T$ is a structured type (tuple, record, class). \\
\hline
\end{tabular}
}%resizebox
\caption{Type operators and relations in \textsc{M}.}
\label{tab:type-ops}
\end{table}
\subsection{Type Coercion and Conversion}
The M language distinguishes between implicit coercion (automatic, safe conversions inserted by the compiler) and explicit conversion (user-directed casts). Implicit coercions ensure smooth interoperability of closely related types without requiring boilerplate, while explicit conversions run the risk of narrowing or lossy transformations and require human awareness. By convention:
\begin{itemize}    
    \item $\hookrightarrow T$ means types that are coercible to $T$ ( $!\hookrightarrow T$ means not), $e:\hookrightarrow T$ means $e$ has type that is coercible to $T$. $any$ is a type placeholder, so $any\hookrightarrow T\ \forall T$.
    \item $\Rightarrow T$ means types that are convertible to $T$ using a default function or user-defined function ($\xRightarrow{f} T$). $e :\Rightarrow T$ is a type casting expression, which produces a new value $e'$ of type $T$, so $val(e:\Rightarrow T)=e', type(e:\Rightarrow T)=T$, 
\end{itemize}
\textbf{Generic Coercion}. If a context $\Theta$ expects parameter $e$ of type $T_2$, while $e$ has type $T1$ and $T_1$ is coercible to $T_2$ then type of $e$ is automatically promoted to $T_2$ in context $\Theta$ (outside $\Theta$, $e$ remains in type $T_1$). If $T_1!\hookrightarrow T_2$, an explicit casting $e:\Rightarrow T_2$ is required.
\begin{mathpar}
\inferrule*[right=generic-coercion]
  { \Gamma \vdash e : T_1,  \Theta(e:T_2),\Gamma \vdash T_1 \hookrightarrow T_2}
  { \Gamma \vdash \Theta[e] : T_2}
\end{mathpar}
\begin{mathpar}
\inferrule*[right=explicit-conversion]
  { \Gamma \vdash e : T_1,  \Theta(e:T_2),\Gamma \vdash T_1 !\hookrightarrow T_2}
  { \Gamma \vdash \Theta[e:\Rightarrow T_2]:T_2}
\end{mathpar}
\\\textbf{Numeric Widening}. $int8 \hookrightarrow int16 \hookrightarrow int32 \hookrightarrow int64$; $uint8 \hookrightarrow uint16 \hookrightarrow uint32 \hookrightarrow uint64$; $int32 \hookrightarrow float$; $int64 \hookrightarrow double$; $float \hookrightarrow double;$
\\\textbf{Boolean, Char, and String}. $char$ implicitly coerces to $string$ of length 1, $bool$ does not implicitly coerce to integers (not like in C language, to avoid ambiguity in modeling semantics, so $var\ i:int = 1 + true;$ will produce a semantic error).
% --- Structured Literals ---
\\\textbf{Tuples and Records}. A tuple is a field-less record, they are interchangeable in contexts if having the same number, order and types of fields.
\begin{mathpar}
\inferrule*[right=coerce-rec2tup]
  { \text{def}~R~\text{record}\{f_1:T_1,\dots,f_n:T_n\}, \text{def}~T~\text{tuple}\{T'_1,\dots,T'_n\},\\\Gamma \vdash T_i\hookrightarrow T'_i~\forall i, e:R, \Theta(e:T)}
  { \Gamma \vdash \Theta[e]:T}
\end{mathpar}
\begin{mathpar}
\inferrule*[right=coerce-tup2rec]
  { \text{def}~T~\text{tuple}\{T_1,\dots,T_n\}, \text{def}~R~\text{tuple}\{f_1:T'_1,\dots,f_n:T'_n\},\\\Gamma \vdash T_i\hookrightarrow T'_i~\forall i, \Gamma \vdash e:T, \Theta(e:R)}
  { \Gamma \vdash \Theta[e]:R}
\end{mathpar}
\\\textbf{Numeric Narrowing}. In contrast to widening, the numeric value is downgraded with possible loss of information (due to rounding or loss of precision).
\begin{mathpar}
\inferrule*[right=conv-numeric]
  { \Gamma \vdash e:T_2, \Gamma \vdash T_1 \hookrightarrow T_2 \hookrightarrow numeric}
  { \Gamma \vdash (e:\Rightarrow T_1) : T_1 }
\end{mathpar}
\\\textbf{Enum and Int}. $int \to enum$: allowed if the value matches one of the enum’s symbolic constants. $enum \to int$: always allowed, yields the underlying integer code.
\begin{mathpar}
\inferrule*[right=conv-int2enum]
  { \Gamma \vdash e:int, \Gamma \vdash E:enum, \exists i: val(E.f_i)=val(e) }
  { \Gamma \vdash e\Rightarrow E : E }
\end{mathpar}
\begin{mathpar}
\inferrule*[right=conv-enum2int]
  { \Gamma \vdash e : E, \Gamma \vdash E: enum }
  { \Gamma \vdash e\Rightarrow int : int }
\end{mathpar}
The handling of $timespan$ type is elaborated in section \ref{sec:timespan}. There are possible semantic errors with type conversions that need to be handled properly, like: multiple coercion paths (ambiguity), narrowing conversions without explicit cast, unresolved coercion when context type is unknown.
\subsection{Type Equality}
\textbf{Type Alias}. Each type is mapped to a unique reference number or ID by the compiler. An alias is a symbolic name that does not create a new type ID, just refers to an existing one. Thus, alias names are interchangeable with their underlying types in typing derivations. Formally, if A is an alias of type T, then any occurrence of A in a typing judgment is equivalent to T:
\begin{mathpar}
\inferrule*[right=alias]
  { \Gamma \vdash e:T, \text{def}\ A\ T}
  { \Gamma \vdash e:A }
\end{mathpar}
\textbf{Contextual Distinction}. In contexts like definition of members in a structured type, comparison of values, types are treated by their nominal names and are only equivalent if they bear the same name.
\begin{mathpar}
\inferrule*[right=context-diff]
  { \Gamma \vdash a:A,\Gamma \vdash b:B,\Theta(a:A,b:B),\Theta \in \{def,compare\}}
  { \Gamma \vdash \Theta[a]:A,\Theta[b]:B}
\end{mathpar}

\subsection{Type Inference for Literals}
\textbf{Scalar Literals}. The syntax of M has been designed to recognize the type of a scalar value from its lexical form.
\begin{mathpar}
\inferrule*[right=literal]
  { \Gamma, expr=v}
  { \Gamma \vdash v:ltype(v) }
\end{mathpar}
\textbf{Collection Literals}. Collection literals coerce to any compatible collection type by element unification:
\begin{mathpar}
\inferrule*[right=coerce-coll]
  { \Gamma \vdash e_i : T_i,\Gamma \vdash T_i \hookrightarrow T_j\ \forall i \land \exists j \in 1..n }
  { \Gamma \vdash C(e_1, \dots, e_n) : C\{T_j,n\} }
\end{mathpar}
where $C$ is a collection constructor ($array, list, set$). If the target type provides a fixed size, the literal size must match. If the target type leaves size unspecified, it is dynamically sized collection. If the element types cannot be unified, it is a type error.
\\\textbf{Tuple Literal}. 
\begin{mathpar}
\inferrule*[right=tupleval-inline]
  { \Gamma  }
  { \Gamma \vdash \text{tuple}(e_1, \dots, e_n) : \text{tuple}\{type(e_1),\dots,type(e_n)\} }
\end{mathpar}
\begin{mathpar}
\inferrule*[right=tupleval-def]
  { \Gamma, \text{def}\ T\ \text{tuple}\{T_1,\dots,T_n\}, \Gamma \vdash e_i :\hookrightarrow T_i, \forall i \in 1..n }
  { \Gamma \vdash \text{T}(e_1, \dots, e_n) : T }
\end{mathpar}
\textbf{Record and Class Literals}.
\begin{mathpar}
\inferrule*[right=recordval]
  { \text{def}\ R\ \text{record}\{f_1:T_1,\dots,f_n:T_n\}, \Gamma \vdash e_i :\hookrightarrow T_i, \forall i \in 1..n }
  { \Gamma \vdash R(f_1:e_1, \dots, f_n:e_n) : R }
\end{mathpar}
\begin{mathpar}
\inferrule*[right=classval]
  { \text{def}\ A\ \text{class}\{f_1:T_1,\dots,f_n:T_n\}, \Gamma \vdash e_i :\hookrightarrow T_i, \forall i \in 1..n }
  { \Gamma \vdash A(f_1:e_1, \dots, f_n:e_n) : A }
\end{mathpar}

\subsection{Typing Rules for Expressions}
\textbf{Assignment}.
\begin{mathpar}
\inferrule*[right=assign]
  { \Gamma \vdash l : T, \Gamma \vdash r:\hookrightarrow T, T!=any}
  { \Gamma \vdash l:T, (l=r):T }
\end{mathpar}
\begin{mathpar}
\inferrule*[right=assign-any]
  { \Gamma \vdash l : any, \Gamma \vdash r: T }
  { \Gamma \vdash l:T, (l=r):T }
\end{mathpar}
\textbf{Extended Assignment}. Like $a\mathrel{{+}{=}b}$ will be expanded into $a = a + b$ and inherit typing rules for assignment and arithmetic expression.
\begin{mathpar}
\inferrule*[right=assign-ext]
  { \Gamma \vdash a:A,\Gamma \vdash b:B}
  { \Gamma \vdash a\mathrel{{+}{=}}b:type(a = a + b) }
\end{mathpar}
\textbf{Conditional Expression}. $a=b?a_1:a_2$, type of condition $b$ must be Boolean, then type of $a$ can be decided by either type of $a_1$ or $a_2$ depending on the value of $b$.
\begin{mathpar}
\inferrule*[right=ifexpr-true]
  { \Gamma \vdash b:bool,val(b)=true}
  { \Gamma \vdash (b?a_1:a_2): type(a_1) }
\end{mathpar}
\begin{mathpar}
\inferrule*[right=ifexpr-false]
  { \Gamma \vdash b:bool,val(b)=false}
  { \Gamma \vdash (b?a_1:a_2): type(a_2) }
\end{mathpar}
\begin{mathpar}
\inferrule*[right=ifexpr-undecided]
  { \Gamma \vdash b:bool,val(b)=null}
  { \Gamma \vdash (b?a_1:a_2): type(a_1) \sqcup type(a_2) }
\end{mathpar}
\textbf{Postfix Expressions}. Including references by membership ($a.b$) or by an indexing key ($a[b]$) or by a call with parameters ($a(v_1,v_2,\dots)$ or $a(k_1:v_1,k_2:v_2,\dots)$).
\begin{mathpar}
\inferrule*[right=member]
  { \Gamma \vdash type(a)\in \{record,class\}, b \in members(a)}
  { \Gamma \vdash (a.b):type(b) }
\end{mathpar}
\begin{mathpar}
\inferrule*[right=member-enum]
  { \Gamma, \text{def}\ E\ \text{enum}\{f_1,\dots,f_n\}}
  { \Gamma \vdash E.f_i:E }
\end{mathpar}
\begin{mathpar}
\inferrule*[right=index-int]
  { \Gamma \vdash a:array|list|set\{T\}, i:int, 0\leq val(i) < length(a)}
  { \Gamma \vdash a[i]:T }
\end{mathpar}
\begin{mathpar}
\inferrule*[right=index-tuple]
  { \Gamma \vdash a:tuple\{T_0,\dots,T_{n-1}\}, i:int, 0\leq val(i)< n}
  { \Gamma \vdash a[i]:T_i }
\end{mathpar}
\begin{mathpar}
\inferrule*[right=index-key]
  { \Gamma \vdash a:map\{T_k,T_v\}, \Gamma \vdash k:\hookrightarrow T_k}
  { \Gamma \vdash a[k]:T_v }
\end{mathpar}
\begin{mathpar}
\inferrule*[right=call-ref]
  { \text{function}\ f(a_1:T_1,\dots,a_n:T_n):T_r, \Gamma \vdash v_i\hookrightarrow T_i \forall i\in 1..n}
  { \Gamma \vdash f(v_1,\dots,v_n):T_r }
\end{mathpar}
\textbf{Arithmetic Expressions}.
\begin{mathpar}
\inferrule*[right=binary-arith1]
  { op \in \{+,-,*\}, \Gamma \vdash type(a)\leadsto type(b)\leadsto numeric}
  { \Gamma \vdash (a\ op\ b): type(b)}
\end{mathpar}
\begin{mathpar}
\inferrule*[right=binary-arith2]
  { op \in \{+,-,*\}, \Gamma \vdash type(b)\leadsto type(a)\leadsto numeric}
  { \Gamma \vdash (a\ op\ b): type(a)}
\end{mathpar}
\begin{mathpar}
\inferrule*[right=binary-div]
  { \Gamma \vdash a:T_1,b:T_2,T_1\leadsto numeric, T_2\leadsto  numeric}
  { \Gamma \vdash (a / b): double}
\end{mathpar}
\begin{mathpar}
\inferrule*[right=binary-mod]
  { \Gamma \vdash a:int,b:int}
  { \Gamma \vdash (a \% b): int}
\end{mathpar}
\begin{mathpar}
\inferrule*[right=unary-arith]
  { \Gamma, op\ a, op \in \{-\}}
  { \Gamma \vdash (op\ a): type(a)}
\end{mathpar}
\textbf{Logical Expressions}.
\begin{mathpar}
\inferrule*[right=binary-logic]
  { \Gamma \vdash a:bool,b:bool,op \in \{\&\&,||\}}
  { \Gamma \vdash (a\ op\ b): bool}
\end{mathpar}
\begin{mathpar}
\inferrule*[right=unary-logic]
  { \Gamma \vdash a:bool,op \in \{!\}}
  { \Gamma \vdash (op\ a): bool}
\end{mathpar}
\textbf{Bitwise Expressions}.
\begin{mathpar}
\inferrule*[right=bitwise-binary]
  { \Gamma \vdash a:int,b:int,op\in\{|,\&,\wedge\}}
  { \Gamma \vdash (a\ op\ b): int}
\end{mathpar}
\begin{mathpar}
\inferrule*[right=bitwise-unary]
  { \Gamma \vdash a:int}
  { \Gamma \vdash (\sim a): int}
\end{mathpar}
\begin{mathpar}
\inferrule*[right=shift]
  { \Gamma \vdash a:int,b:uint,op\in\{<<,>>\}}
  { \Gamma \vdash (a\ op\ b): int}
\end{mathpar}
\textbf{Comparison Expressions}.
\begin{mathpar}
\inferrule*[right=cmp-eq]
  { \Gamma \vdash a:T,b:T,T!=any,op \in \{==,!=\}}
  { \Gamma \vdash (a\ op\ b): bool}
\end{mathpar}
\begin{mathpar}
\inferrule*[right=cmp-rel]
  { \Gamma \vdash a:\leadsto T,b:\leadsto T,T \in \{numeric,timespan\},op \in \{<,<=,>,>=\}}
  { \Gamma \vdash (a\ op\ b): bool, (a\ op\ b) \equiv (a:\Rightarrow T\ op\ b:\Rightarrow T)}
\end{mathpar}

\subsection{Typing Rules for Statements}
\textbf{Variable and Constant Declarations}.
\begin{mathpar}
\inferrule*[right=var-decl]
  { \Gamma,var|const\ x:T=v;}
  { \Gamma \vdash x:type(x=v) }
\end{mathpar}
\begin{mathpar}
\inferrule*[right=var-decl-novalue]
  { \Gamma,var\ x:T;}
  { \Gamma \vdash x:T }
\end{mathpar}
\textbf{Sequence Statement}. Take the last non-any type if possible
\begin{mathpar}
\inferrule*[right=seq-any]
  { \Gamma \vdash s_2:any}
  { \Gamma \vdash (s_1;s_2):type(s_1)}
\end{mathpar}
\begin{mathpar}
\inferrule*[right=seq-nonany]
  { \Gamma \vdash type(s_2)!=any}
  { \Gamma \vdash (s_1;s_2):type(s_2)}
\end{mathpar}
\textbf{Return Statement}. Returns the type of the returned value and ignore subsequent statements.
\begin{mathpar}
\inferrule*[right=return]
  { \Gamma \vdash e:T}
  { \Gamma \vdash (\text{return}\ e;s*;):T}
\end{mathpar}
\begin{mathpar}
\inferrule*[right=return-any]
  { \Gamma }
  { \Gamma \vdash (\text{return};s*;):any}
\end{mathpar}
\textbf{Break Statement}. Returns the type before the break point.
\begin{mathpar}
\inferrule*[right=break-any]
  { \Gamma }
  { \Gamma \vdash (s_1*;\text{break};s_2*;):type(s_1*)}
\end{mathpar}
So combining the sequence, return and break rules, the type of a chain of statements $s*$ or a block $\{s*\}$ is decidable.
\\\\\textbf{Function Declaration}.
\begin{mathpar}
\inferrule*[right=func-decl]
  { \text{function}\ f(a_1:T_1,\dots,a_n:T_n):T_r\ block, \Gamma \vdash block:\hookrightarrow T_r}
  { \Gamma \vdash f:T_r}
\end{mathpar}
\textbf{Do Declaration}.
\begin{mathpar}
\inferrule*[right=do-every]
  { \text{do[every(t)]}, \Gamma \vdash t:\hookrightarrow timespan}
  { \Gamma \vdash}
\end{mathpar}
\begin{mathpar}
\inferrule*[right=do-on]
  { \text{do}[\text{on}(c)],\Gamma \vdash c:\hookrightarrow bool}
  { \Gamma \vdash}
\end{mathpar}%
\begin{mathpar}
\inferrule*[right=do-recv]
  { \Gamma, \text{do}~ID(a_1:T_1,\dots,a_n:T_n)}
  { \Gamma \vdash}
\end{mathpar}
\textbf{If Statement}.
\begin{mathpar}
\inferrule*[right=if-true]
  { \Gamma \vdash b:bool, val(b)=true}
  { \Gamma \vdash (\text{if}\ (b)\ block_1\ (\text{else}\ block_2)?):type(block_1) }
\end{mathpar}
\begin{mathpar}
\inferrule*[right=if-false]
  { \Gamma \vdash b:bool, val(b)=false}
  { \Gamma \vdash (\text{if}\ (b)\ block_1):any }
\end{mathpar}
\begin{mathpar}
\inferrule*[right=ifelse-false]
  { \Gamma \vdash b:bool, val(b)=false}
  { \Gamma \vdash (\text{if}\ (b)\ block_1\ \text{else}\ block_2):type(block_2) }
\end{mathpar}
\begin{mathpar}
\inferrule*[right=if-undecided]
  { \Gamma \vdash b:bool, val(b)=null}
  { \Gamma \vdash (\text{if}\ (b)\ block_1\ (\text{else}\ block_2)?):any }
\end{mathpar}
\textbf{Case Statement}. Types of all options must be coercible to the type of the selector.
\begin{mathpar}
\inferrule*[right=cases-one]
  { \Gamma \vdash e:T,v_i:T_i,T_i\hookrightarrow T \forall i \in 1..n,\exists j\in 1..n:val(e)=val(v_j)}
  { \Gamma \vdash (\text{cases}(e)\{\text{case}\ v_1: s_1*;\dots \text{case}\ v_n: s_n*; \text{otherwise}: o*;\}):type(s_j*) }
\end{mathpar}
\begin{mathpar}
\inferrule*[right=cases-other]
  { \Gamma \vdash e:T,v_i:T_i,T_i\hookrightarrow T \forall i \in 1..n,val(e)!=val(v_i)\forall i}
  { \Gamma \vdash (\text{cases}(e)\{\text{case}\ v_1: s_1*;\dots \text{case}\ v_n: s_n*; \text{otherwise}: o*;\}):type(o*) }
\end{mathpar}
\textbf{While Statement}.
\begin{mathpar}
\inferrule*[right=while-true]
  { \Gamma \vdash b:bool,val(b)=true,block:T}
  { \Gamma \vdash (\text{while}\ (b)\ block):T }
\end{mathpar}
\begin{mathpar}
\inferrule*[right=while-nottrue]
  { \Gamma \vdash b:bool,val(b)!=true,block:T}
  { \Gamma \vdash (\text{while}\ (b)\ block):any }
\end{mathpar}
\textbf{Foreach Statement}.
\begin{mathpar}
\inferrule*[right=foreach-keys]
  { \Gamma \vdash a\in\{array,list,set\},k:int}
  { \Gamma \vdash (\text{foreach}\ (\text{var}\ k\ \text{in}\ keys(a))\ block):type(block) }
\end{mathpar}
\begin{mathpar}
\inferrule*[right=foreach-values]
  { \Gamma \vdash a\in\{array,list,set,map\},v:type(elem(a)}
  { \Gamma \vdash (\text{foreach}\ (\text{var}\ v\ \text{in}\ values(a))\ block):type(block) }
\end{mathpar}
\begin{mathpar}
\inferrule*[right=foreach-pairs]
  { \Gamma \vdash a:map\{T_k,T_v\},k:T_k,v:T_v}
  { \Gamma \vdash (\text{foreach}\ (\text{var}\ k,v\ \text{in}\ pairs(a))\ block):type(block) }
\end{mathpar}
\textbf{Tell Statement}. A tell statement returns true or false, necessary type checks are: recipient actor address, task name, content parameters and extra parameters in $with$ clause.
\begin{mathpar}
\inferrule*[right=tell]
  { \Gamma \vdash r:A, A=\text{class}\{\dots \text{do}\ \text{doSmth}\{fa_1:T_1,\dots,fa_n:T_n\}\dots\},
  \\a_1:\hookrightarrow T_1,\dots,a_n:\hookrightarrow T_n,
  \\w_{a}:\text{timespan},w_{d}:\text{timespan},w_{s}:\text{actor},
  }
  { \Gamma \vdash r!\text{doSmth}(a_1,\dots,a_n)\: \text{with}(\text{after}:w_{a},\text{deadline}:w_{d},\text{sender}:w_{s}):bool }
\end{mathpar}
\textbf{Cancel Statement}.
\begin{mathpar}
\inferrule*[right=cancel-all]
  { \Gamma }
  { \Gamma \vdash (\text{cancel}\ *):bool}  
\end{mathpar}
\begin{mathpar}
\inferrule*[right=cancel-names]
  { \Gamma \vdash ev_i:ID|string\ \forall i\in 1..n}
  { \Gamma \vdash (\text{cancel}\ ev_1,\dots,ev_n):bool}  
\end{mathpar}
\subsubsection{Semantic Rules for Timespan Type}\label{sec:timespan}
M uses a \texttt{SIM\_TIME\_UNIT} constant (formalized as $\mu_{\hat{\tau}}$) to define the duration represented by one logical clock unit and the compiler will first convert all timespan literals to miliseconds as a common base, and then to this unit. So let $\mu_{\hat{\tau}}$ be a $timespan$ constant, e.g. $1ms$. Let $scale(u)$ be the number of $\mu_{\hat{\tau}}$ units equal to one unit of $u$, so given $\mu_{\hat{\tau}}=1ms$ then $scale(1s)=1000,scale(1ms)=1$. A timespan literal $d$ is expressed as the sum of components with explicit units: 
\[
d=n_1u_1 + n_2u_2 + \dots\ \text{where}\ n_i:\text{double}, u_i\in\{ns,us,ms,s,m,h,d\}
\]
% Conversion function
\newcommand{\timeval}[1]{\llbracket #1 \rrbracket_{\hat{\tau}}}
Let $\timeval{d}$ be the function to convert a $timespan$ $d$ to a number of $\mu_{\hat{\tau}}$ units:
\[
\timeval{d} = \sum_{i} n_i.scale(u_i)
\]
For a numeric value $n$, $\timeval{n}=n$ in $\mu_{\hat{\tau}}$ units.
Then we have following rules for converting and comparing timespans:
% Timespan typing and conversion
\[
\frac{ \Gamma \vdash d : \text{timespan},\text{const}\ \mu_{\hat{\tau}}:\text{timespan}  }
     { \Gamma \vdash val(d:\Rightarrow \text{double})=\timeval{d} (\mu_{\hat{\tau}}\ \text{units}) }
     \quad\textsc{(timespan-to-num)}
\]

% Numeric literals as implicit timespans
\[
\frac{ \Gamma \vdash n : \text{numeric},\text{const}\ \mu_{\hat{\tau}}:\text{timespan} }
     { \Gamma \vdash val(n:\Rightarrow \text{timespan})=n (\mu_{\hat{\tau}}\ \text{units})
     }
     \quad\textsc{(num-to-timespan)}
\]

% Comparison rule
\[
\frac{ \Gamma \vdash d_1:\text{timespan} 
       \quad 
       \Gamma \vdash d_2:\text{timespan}} 
     { d_1 \;\bowtie\; d_2 
       \;\equiv\; 
       \timeval{d_1} \;\bowtie\; \timeval{d_2} }
     \bowtie \in\{<,\leq,==,!=,\geq,>\}  
     \quad\textsc{(cmp-timespan)}
\]
\subsection{Construction of Compiler and Generators}
We are building an M compiler that enables easy creation of custom generators. The chosen technology stack is as follows:
\begin{itemize}    
    \item ANTLR4 \cite{10.5555/2501720}\footnote{\url{https://www.antlr.org/}} is chosen as the parser generation tool.
    \item JastAdd \cite{hedin09}\footnote{\url{https://jastadd.cs.lth.se/}} is used to define the AST structure and semantic attributes, and do semantic analysis in a declarative, modular way.
    \item Java is the main programming language as ANTLR4 and JastAdd are based on and integrate well with Java.
    \item Code generators are Java-based applications that utilize the compiler's decorated AST to generate different target codes. Users just need to inherit from a highly reusable base generator and customize with their own specifics.
\end{itemize}
There are exemplary compilers that are built similarly for languages like Java \cite{10.1145/1297105.1297029}, Modelica \cite{AKESSON2008117}, or ABS \cite{10.1007/978-3-642-25271-6_8}. The chosen technical approach is therefore viable and maintainable. Of course, it is only one way of building a compiler, there are many other tools and languages for parser generation \cite{parsers} and semantics specification and analysis (e.g. RML \cite{5076648}, K framework \cite{10.1007/978-3-030-55089-9_4}, Spoofax \cite{DBLP:conf/birthday/Kalleberg23}, Rascal \cite{8930879}). We only head for a prototype compiler in this stage, improvements or even re-engineering can be done later in a separate follow-up project. 

\section{Transformability} \label{sec:trans}
In this section we demonstrate the conversion between M and several formalisms and languages. The example codes are of course written manually, conceptually, following the semantics and syntax of M, since at this stage the M compiler is still under development. We can see that the high-level and low-level expressiveness of M are powerful enough to simulate other formalisms, and the codes form patterns that can be automatized through a dedicated code generator for each mapping scheme.
\subsection{State Machine in M}
M itself forms a timed transition system. States, guards and transitions in a state machine can be easily expressed in M:
\begin{itemize}    
    \item State: the state of a machine can be represented as a combination of state variables ($var[state]$) or the user can define a record type with multiple fields and define a single state variable of this type.
    \item On entering or leaving a state: M captures this change by checking the previous and current values and firing a corresponding event when it happens. 
    \begin{lstlisting}[style=CodeBlock]
    do[on(prev(s)!=s1 && s==s1]) enterStateS1{...}
    do[on(prev(s)==s1 && s!=s1]) leaveStateS1{...}
    \end{lstlisting}
    \item On state timeout: schedule a state change event after the timeout period with the $tell$ command and the $with(after:timeout)$ clause.
    \begin{lstlisting}[style=CodeBlock]
        //define the state expiry event
        do setNextState(next:state_type){
            self.s = next;
        }

        //somewhere
        s = s1;
        self!setNextState(s2) with(after:1s);
    \end{lstlisting}
    
    \item On some logical condition: the $do[on(condition)]\{...\}$ construct of M helps define straightly conditional guards and actions.
\end{itemize}

\begin{figure}[H]
    \centering
    \includegraphics[width=0.7\textwidth]{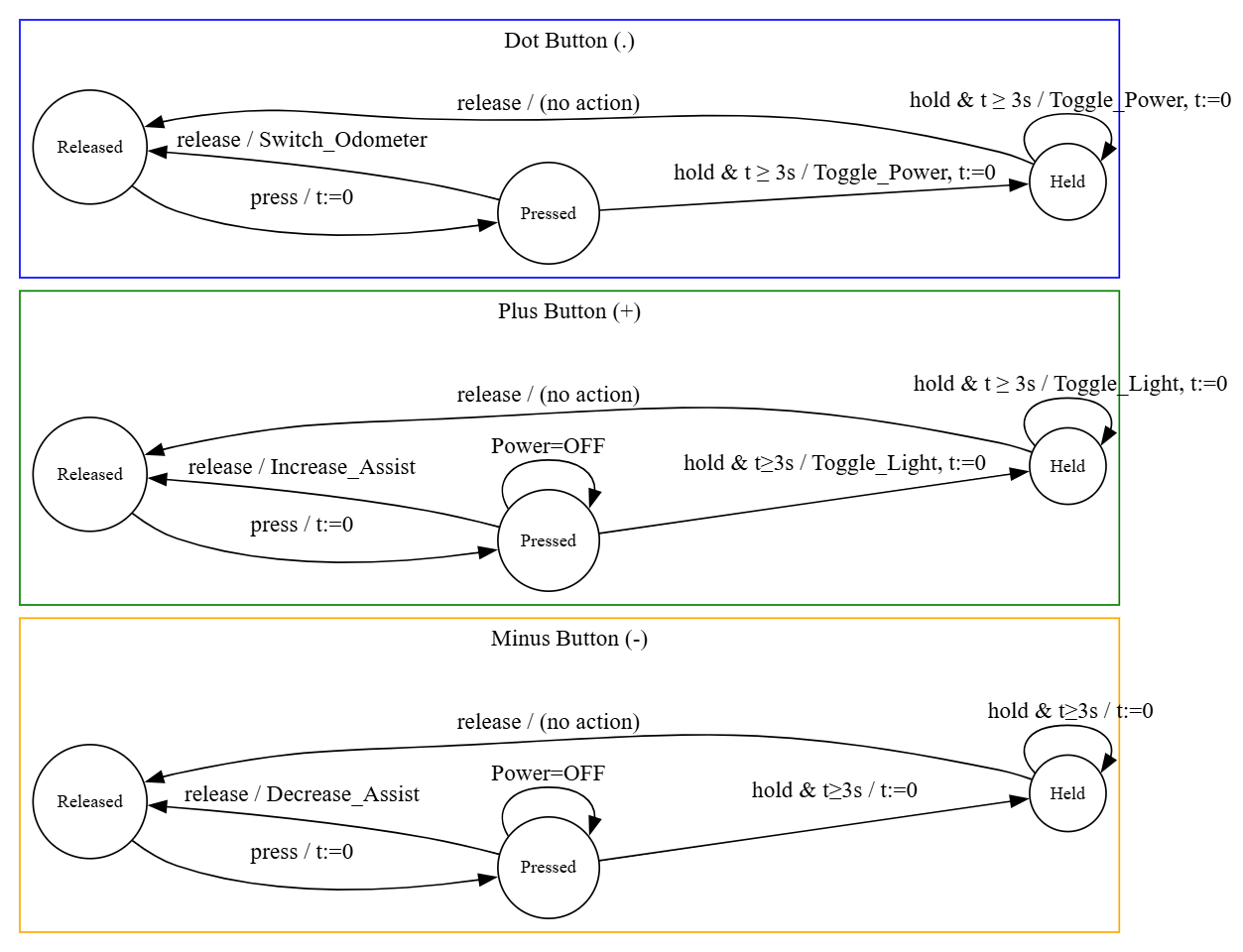}
    \caption{Electric bicycle control panel}
    \label{fig:stm}
\end{figure}

Example \ref{lst:stm}: An electric bicyle's control panel has 3 buttons: $(.)$, $(+)$ and $(-)$. Each button has a short press mode (press and release after $<3$ seconds) and long press mode (press and hold $\geq3$ seconds).
\begin{itemize}
    \item Button $(.)$: long press switches power ON and OFF. Short press switches between 3 display modes of odometer when power is ON.
    \item Button $(+)$: only active when power is ON. Long press switches light ON and OFF. Short press circulates assistance level upwards (0-1-2-3-0 ...).
    \item Button $(-)$: only active when power is ON. Short press circulates assistance level downwards (3-2-1-0-3 ...).
\end{itemize}
The control logic is visualized as 3-region state machine in Fig.\ref{fig:stm}.
\begin{multicols}{2}
    \begin{lstlisting}[label=lst:stm,style=CodeBlock, caption={State machine in M}]
const SIM_TIMEUNIT:timespan = 1ms;
def time double;
def ButtonStates enum{ Released, Pressed, Held };
def ControlPanel class[actor] {

    // Local state variables
    var[state] Power:bool = false;
    var[state] OMode:int = 0;//odometer modes
    var[state] AssistLevel:int = 0;//0..3
    var[state] Light:bool = false;

    // Button state and hold time
    var[state] DotState:ButtonStates = ButtonStates.Released;
    var[state] PlusState:ButtonStates = ButtonStates.Released;
    var[state] MinusState:ButtonStates = ButtonStates.Released;

    //User press button (.)
    do pressDot {
        DotState = ButtonStates.Pressed;
        //plan the long press
        self!longDot() with(after:3s);
    }
    //User release button (.)
    do releaseDot {
        cancel longDot;
        if (Power && DotState == ButtonStates.Pressed){
            OMode = (OMode + 1) % 3;
        }
        DotState = ButtonStates.Released;
    }
    //Button (.) on long press
    do longDot(){
        DotState = ButtonStates.Held;
        Power = !Power;
        //plan the next long press
        self!longDot() with(after:3s);
    }

    do pressPlus { 
        PlusState = ButtonStates.Pressed;
        self!longPlus() with(after:3s);        
    }

    do releasePlus{
        cancel longPlus;
        if (Power && PlusState == ButtonStates.Pressed){
            AssistLevel++;
            if (AssistLevel>3) AssistLevel = 0;
        }
        PlusState = ButtonStates.Released;
    }

    do longPlus {
        PlusState = ButtonStates.Held;
        Light = !Light;
        self!longPlus() with(after:3s);
    }

    do pressMinus {
        MinusState = ButtonStates.Pressed;
        self!longMinus() with(after:3s);        
    }

    do releaseMinus {
        cancel longMinus;
        if (Power && MinusState == ButtonStates.Pressed){
            AssistLevel--;
            if (AssistLevel<0) AssistLevel = 3;
        }
        MinusState = ButtonStates.Released;
    }

    do longMinus{
        MinusState = ButtonStates.Held;
        self!longMinus() with(after:3s);
    }
}        
    \end{lstlisting}
\end{multicols}

\subsection{Parallel DEVS to M}
An atomic model in Parallel DEVS can be simulated by an actor in M:
\begin{enumerate}
    \item The output generation function $\lambda$: equal to an action at the end of a state's lifetime, before switching to a new state.
    \item The $\delta_{ext}$ function: equal to different $do$'s defining responses to different types of incoming messages.
    \item The $\delta_{int}$ function: equal to internal events scheduled to happen at the end of each state's lifetime.
    \item The $ta$ function: equal to a class method in M for dynamically computing the lifetime of a given state.
     \item The $\delta_{conf}$ function: equal to the tie-breaking function in M, and by default, is abstracted from the developer.    
\end{enumerate}
Parallel DEVS is a subset of M as M can simulate any Parallel DEVS model, but not the other way around. The M code in example \ref{lst:pdevs} below realizes the 'interuptable traffic light' model introduced in \cite{TendelooV18}.
\begin{paracol}{2}
\begin{leftcolumn}
\begin{lstlisting}[style=CodeBlock,caption=Parallel DEVS model]   
\end{lstlisting}
\begin{figure}[H]
    \centering
    \includegraphics[width=0.8\linewidth]{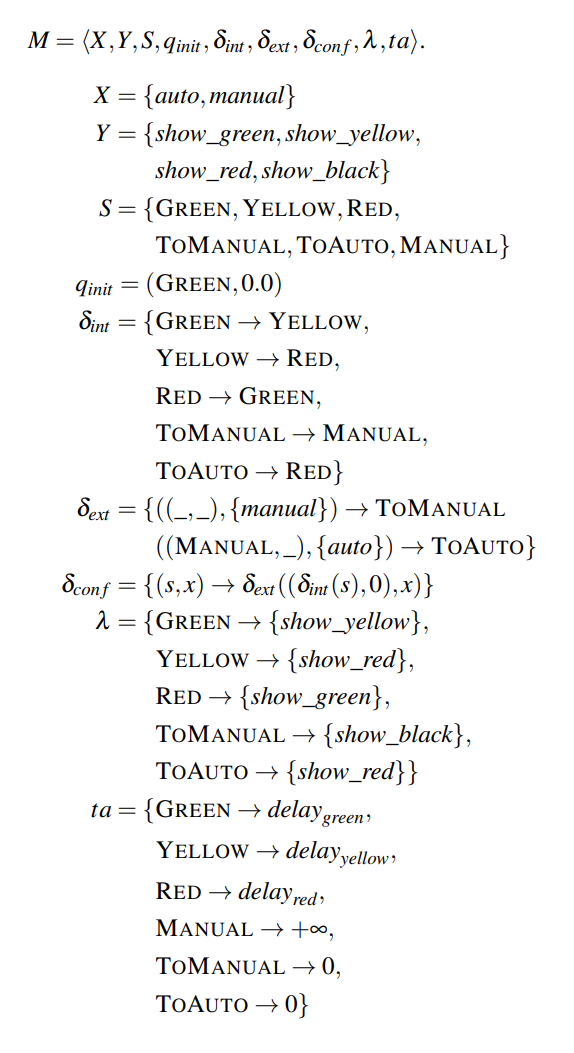}
    %\caption{Parallel DEVS model}
    \label{fig:pdevs-model}
\end{figure}
\end{leftcolumn}
\begin{rightcolumn}
    \begin{lstlisting}[label=lst:pdevs,style=CodeBlock, caption={Parallel DEVS to M}]

const SIM_TIMEUNIT:timespan = 1s;

def States enum{
    UNDEFINED,GREEN,YELLOW,RED,TOMANUAL,TOAUTO,MANUAL
};

def TrafficLight class[actor]{
    var[state] color:States = States.UNDEFINED; 
    var lasttime:real;

    const delay_red:timespan = 30s;
    const delay_yellow:timespan = 5s;
    const delay_green:timespan = 30s;
    const delay_inf:timespan = 1000d;
    
    //PDEVS q_init
    do initialize{        
        self!setState(States.GREEN);
    }

    //PDEVS delta_ext when input=manual
    do ext_manual{
        self!setState(States.TOMANUAL);
    }

    //PDEVS delta_ext when input=auto
    do ext_auto{
        if (color == Modes.MANUAL){
            self!setState(States.TOAUTO);
        }
    }

    //PDEVS internal transition
    do setState(next:States){
        genOutput(); //generate output upto the current state
        self.color = next;
        lasttime = now();
        self!setState(getNextState(self.color)) with(after:ta(self.color));
    }      

    //PDEVS delta_int: get next state of a given state
    function getNextState(s:States):States{
        return cases(s){
            case States.GREEN: States.YELLOW;
            case States.YELLOW: States.RED;
            case States.RED: States.GREEN;            
            case States.TOMANUAL: States.MANUAL;
            case States.TOAUTO:States.RED;
        }
    }

    function ta(c:States):timespan{
        return cases(c){
            case States.RED: delay_red;
            case States.YELLOW: delay_yellow;
            case States.GREEN: delay_green;
            case States.MANUAL: delay_inf;
            case States.TOMANUAL: 0s;
            case States.TOAUTO: 0s;
            otherwise: 0s;
        }
    }

    //PDEVS lamda: generate output upto current state
    function genOutput(){
        cases(self.color){
            case States.GREEN: println("show_yellow");
            case States.YELLOW: println("show_red");
            case States.RED: println("show_green");
            case States.TOMANUAL: println("show_black");
            case States.TOAUTO: println("show_red");
            otherwise:;
        }
    }    

    function elapsed():real{ 
        return now() - self.lasttime; 
    }
}
    \end{lstlisting}
\end{rightcolumn}
\end{paracol}

\subsection{M to ROS2}
ROS2 is a popular platform for developing robotic software in research. We can model robotic systems with node graph architecture in M and generate ROS2 implementation code therefrom. M maps directly to ROS2 constructs:
\begin{itemize}
    \item Nodes: an actor class in M maps to a node class in ROS2, a $do$ maps to a $callback$, $do$'s formal arguments map to fields in ROS2's corresponding message format.
    \item Publish-subscribe channels: a connection class in M represents a channel, its properties help to define the corresponding channel in ROS (name, different parameters) and related actions like register as publisher or subscriber, publish and listen.
    \item Environmental constants map to launch parameters in ROS2.
\end{itemize}

Example \ref{lst:ros2} shows how the minimal publisher and subscriber nodes are expressed in M and ROS2 (\href{https://docs.ros.org/en/kilted/Tutorials/Beginner-Client-Libraries/Writing-A-Simple-Cpp-Publisher-And-Subscriber.html}{"Kilted"} version).
\begin{paracol}{2}
\begin{leftcolumn}   
\begin{lstlisting}[label=lst:ros2,style=CodeBlock, caption={ROS2 nodes in M}]
//file connection.m
def PubSubChannel class[connection]{
    var name:string;
    var publishers:map{actor,bool};
    var subscribers:map{actor,string};

    do register_pub(me:actor){
        publishers[me]=true;        
    }

    do register_sub(me:actor,callback:string){
        subscribers[me]=callback;        
    }

    do publish(string m){
        foreach(var listener:actor,callback:string in subscribers.entries(){ 
                listener!callback(m);
        }
    }
};

//file publisher.m
include "connection.m";
def MinimalPublisher class[actor]{
    var name:string;
    var count:int=0;
    var channel:connection;
    
    do initialize{
        channel!register_pub(self);
    }
    
    do[every(500ms)] timer_callback{
        count++;
        channel!publish(string.format("Hello, world! %d",count));
    }
};
main{
    var ch:connection = PubSubChannel(name:"topic");
    var publisher:actor = MinimalPublisher(name:"minimal_publisher",channel:ch);
}

//file subscriber.m
include "connection.m";
def MinimalSubscriber class[actor]{
    var name:string;
    var channel:connection;

    do initialize{
        channel!register_sub(self,"topic_callback");
    }

    do topic_callback(string m){
        println(string.format("I heard: '%s'",m));
    }
};   
main{
    var channel:connection = PubSubChannel(name:"topic");
    var subscriber:actor = MinimalSubscriber(name:"minimal_subscriber",channel:ch);
}
\end{lstlisting}
\end{leftcolumn}
\begin{rightcolumn}
\begin{lstlisting}[style=CodeBlock, caption={ROS2 nodes in C++}]
//file publisher.cpp
class MinimalPublisher : public rclcpp::Node {
public:
  MinimalPublisher() : Node("minimal_publisher"), count_(0) {
    publisher_ = this->create_publisher<std_msgs::msg::String>("topic", 10);
    auto timer_callback =
      [this]() -> void {
        auto message = std_msgs::msg::String();
        message.data = "Hello, world! " + std::to_string(this->count_++);
        RCLCPP_INFO(this->get_logger(), "Publishing: '%s'", message.data.c_str());
        this->publisher_->publish(message);
      };
    timer_ = this->create_wall_timer(500ms, timer_callback);
  }

private:
  rclcpp::TimerBase::SharedPtr timer_;
  rclcpp::Publisher<std_msgs::msg::String>::SharedPtr publisher_;
  size_t count_;
};

//file subscriber.cpp
class MinimalSubscriber : public rclcpp::Node {
public:
  MinimalSubscriber() : Node("minimal_subscriber") {
    auto topic_callback =
      [this](std_msgs::msg::String::UniquePtr msg) -> void {
        RCLCPP_INFO(this->get_logger(), "I heard: '%s'", msg->data.c_str());
      };
    subscription_ =
      this->create_subscription<std_msgs::msg::String>("topic", 10, topic_callback);
  }

private:
  rclcpp::Subscription<std_msgs::msg::String>::SharedPtr subscription_;
};
\end{lstlisting}   
\end{rightcolumn}
\end{paracol}
\subsection{M to Timed Rebeca}
We intend to generate Timed Rebeca models from M for model-checking. Since Timed Rebeca operates with discrete states and transitions, the conversion requires discretizing continuous state variables and transaction parameters, as discussed in \cite{ros2rebeca_journal}. 

Example~\ref{lst:m2rebec} illustrates the process with the classical \textit{bouncing ball} scenario where a ball is dropped from an initial height, falls under gravity, bounces on the ground, and gradually comes to rest depending on the restitution coefficient $r$. Physically, the motion is continuous, governed by differential equations of height $y$, velocity $v$ and time $t$: $dv/dt = -9.81,\ dy/dt = v$. Since M does not yet support direct modeling of continuous dynamics, we approximate the behavior by updating position and velocity discretely every micro timestep $dt$. 

\textbf{In M}, the ball is represented as an actor class with state variables \texttt{y} and \texttt{v}. A periodic action \texttt{do[every(dt)]} updates the state according to Newtonian mechanics and resets velocity upon bouncing. 

\textbf{In Timed Rebeca},  the same model is mapped to a \texttt{reactiveclass}, with the periodic action realized by self-scheduled messages using \texttt{after(dt)}. Continuous variables marked with \emph{discretize} are disretized to integers through extra functions.

\begin{paracol}{2}
\begin{leftcolumn}
\begin{lstlisting}[label=lst:m2rebec,style=CodeBlock, caption={Bouncing ball model in M}]
const g:double = 9.81;
const r:double = 0.9;
const SIM_TIME_UNIT:timespan = 1000ms;
const period:timespan = 10ms;
const initY:double = 10.0;

def Ball class[actor] {
    var[state,discretize] y:double;
    var[state,discretize] v:double;
    
    do[every(period)] update{
        v = v - g * dt();
        y = y + v * dt();
        if (y < 0) {
            y = 0;
            v = -r * v;
        }
    }
    //dt is a fraction of a second
    function dt():double{
        return (period as double) 
            / (SIM_TIME_UNIT as double);
    }
};

main {
    Ball b(y:initY, v:0.0);
}
\end{lstlisting}   
\end{leftcolumn}
\begin{rightcolumn}
\begin{lstlisting}[style=CodeBlock, caption={Timed Rebeca model}]
//Write your code here !
env double g = 9.81;
env double r = 0.9;
env int SIM_TIME_UNIT = 1000;
env int period = 10;
env double initY = 10.0;

reactiveclass Ball(2) {
    statevars {
        int y_;
        int v_;
        int initY_; //to facilitate LTL checks
    }

    Ball(double y, double v) {
        y_ = discretize(y); 
        v_ = discretize(v); 
        initY_ = y_;
        self.update() after(period);
    }

    msgsrv update() {
        double y = undiscretize(self.y_);
        double v = undiscretize(self.v_);
        
        v = v - g * dt();
        y = y + v * dt();
        if (y < 0) {
            y = 0;
            v = -r * v;
        }
        
        self.y_ = discretize(y);
        self.v_ = discretize(v);
        
        self.update() after(period);
    }

    double dt(){
      return (double)(period)/(double)SIM_TIME_UNIT;
    }
    
    int discretize(double x){
      return (int)(x*100000.0);
    }

    double undiscretize(int x){
      return (double)(x)/1000000.0;
    }
}

main {
    Ball ball():(initY,0.0);
}
\end{lstlisting}   
\end{rightcolumn}
\end{paracol}

\textbf{Verifiable Properties}. Once expressed in Timed Rebeca, the model can be analyzed with its model checker. Interesting properties to verify may include\footnote{Due to discrete approximation, it may require some tolerance around 0, but we do not go into more details here.}:
\begin{itemize}
    \item \textbf{Reality}. The ball never goes below ground: $\Box (y \geq 0)$
    \item \textbf{Energy loss}. In general $r\leq 1$ so the ball loses motion energy after each bounce and does not reach the original height: $\Box( y \leq initY )$
    \item \textbf{Eventual rest (for $r<1$)}. Eventually, the ball comes to rest on the ground: $\Diamond (y = 0 \wedge v = 0)$
    \item \textbf{Recurrent bouncing (for $r=1$)}. Every time the ball hits the ground, it will bounce again: $\Box (y=0 \rightarrow \Diamond (y > 0))$
\end{itemize}
The properties are expressed as below in Rebeca's \href{https://rebeca-lang.org/assets/documents/Handbook-of-Rebeca.pdf}{LTL notations}:
\begin{lstlisting}[style=CodeBlock, caption={Safety properties in Rebeca}]
property{
    define{
        r_normal=r<=1.0; r_one=r==1.0; r_less=r<1.0;
        above=y_>0; onground=y_==0; lower=y_<=initY_;
        stopped=v_==0;
    }
    LTL{
        reality: G(above||onground);
        energy_loss: G(r_normal) && G(lower);
        rest: G(r_less) && F(onground && stopped);
        bounce: G(r_one) && G(onground -> F(above));
    }
}  
\end{lstlisting}
This example demonstrates how an M specification can be systematically converted to a formal model for verification, bridging high-level modeling and rigorous analysis. We will have to workout mechanisms to declare different discretization schemes, using an extra property as in the example is only one possibility.

\section{Comparison} \label{sec:cmp}
%In this section we compare M with several modeling/simulation languages and platforms to mainly reason why we have to create a new language instead of using an existing one. The comparisons are of course not exhaustive and empirical, but to highlight the major reasons that discourage us from selecting them as M and motivate the design of M. Most of them are:
%\begin{enumerate}
%\item The language is not compilable because there is no clearly defined semantics behind.
%\item The compiler was not designed for multi-target compilation. If we want to use the language as M, we have to rework the whole compiler which in many cases is too complicated.
%\item The language itself is too complicated for transforming to other formalisms. As it was designed for one single purpose like simulation or implementation, that complication can be handled, but for multiple targets we have to think about reducing complexity and seeking equivalence. 
%\item The language does not include a time semantics for modeling time-aware behaviours.
%\item The language contains semantic-shaded areas like implementation-oriented codes, external codes that are not easily deduce semantics out of such. 
%\end{enumerate}

In this section we compare \textsc{M} with several existing modeling and simulation languages and platforms, with the primary goal of explaining why we chose to design a new language rather than adopt an existing one. The discussion is not intended to be exhaustive or empirical; rather, it highlights the key factors that discourage us from selecting these languages as the foundation for \textsc{M} and that motivate our design choices. The main reasons are:
\begin{enumerate}
\item \textbf{Lack of formal semantics}. Many languages are not truly compilable because their semantics are not precisely defined. This makes it difficult to reason about correctness and to support systematic compilation.

\item \textbf{Limited compiler architecture}. Existing compilers are often tied to a single execution platform and were not designed for multi-target compilation. Re-engineering such compilers to serve as the backend of \textsc{M} would in many cases be more complex than starting from scratch.

\item \textbf{Overly complex language constructs}. Several languages are designed with one primary use case in mind (e.g., simulation or implementation). While this complexity can be acceptable within that scope, it becomes a barrier when targeting multiple formalisms. For \textsc{M}, we aim to minimize complexity and seek semantic equivalence across backends.

\item \textbf{Insufficient time semantics}. Some candidate languages lack explicit support for modeling time-aware behaviors. Since time is fundamental to modeling dynamic and concurrent systems, e.g. as argued by Lee in \cite{Lee2018ModelsOfTimedSystems}, this omission is critical.%~\cite{DBLP:journals/sosym/Selic12,DBLP:journals/jzusc/LiuZCLGW20}

\item \textbf{Semantic ambiguities}. Some languages contain areas with weakly defined or shaded semantics, such as implementation-specific constructs or foreign code blocks. These are problematic when the goal is to provide a clear and analyzable semantic foundation.
\end{enumerate}

\subsection{M vs. UML}
UML is a general-purpose, largely graphical notation whose metamodel captures structure well but leaves precise execution semantics to auxiliary means (e.g., action languages such as Alf or "opaque actions" in SysML v2~\cite{umlalf,sysmlv2}). This separation has long been criticized for enabling multiple, incompatible interpretations of behavior and for deepening semantic gaps during transformation to simulators or code generators~\cite{lund2010}. As a result, UML models are typically not executable "by construction", and significant, tool-specific effort is required to operationalize them for simulation, deployment, or verification. M addresses these limitations by being a textual, grammar-driven language with an explicit operational semantics: actor-based concurrency combined with DE time and scheduling, state variables, event scanning, and queue manipulation (e.g., $with(after: …)$, $cancel$). This makes M models directly executable/simulatable and amenable to semantics-preserving compilation. 

In the M toolchain, parsing and semantic analysis are performed once; the resulting decorated AST is then reused across multiple generators to produce: (i) simulation artifacts, (ii) inputs to domain-specific compilers and code generators, (iii) native executables via MLIR/LLVM~\cite{mlir21,llvm}, and (iv) models for formal verification. This multi-target capability was a first-class design goal of M and its tooling. Practically, UML models can still serve as sources in the M toolchain; textualization (e.g., via PlantUML as an intermediate representation) allows mapping UML to M, after which all M’s back-ends become available. In this role, M acts as a semantics-aware pivot: once a UML-to-M adapter captures the intended behaviors, the same M model can be simulated, verified, and compiled toward multiple platforms without re-writing behavioral semantics for each target. This reduces semantic discontinuities while improving reuse across the life-cycle.

\subsection{M vs. Metamodel-based Languages}
Metamodel-based approaches (e.g., EMF/Ecore within the OMG MOF stack)~\cite{DBLP:journals/sosym/Brown04,karma1,karma2,karma3} excel at structural definition and at organizing model-to-model/text transformations. However, the metamodel typically does not contain operational semantics; executable behavior must be attached later (e.g., through transformation rules, action languages, or embedded code). While patterns such as using an intermediate metamodel help curb the N×M explosion of pairwise transformations, the absence of first-class, agreed execution semantics often complicates faithful simulation and multi-platform code generation~\cite{DBLP:journals/jss/LanoRTS18}. M complements and can anchor such ecosystems by providing a concrete, executable semantics (actors + DE scheduling) in a textual language engineered with compiler principles. In the M toolchain, DSLx-to-M adaptors encode the meaning of a domain-specific metamodel once, after which the same M core can be compiled to multiple targets (simulation platforms, verification formalisms, MLIR/LLVM, or platform-specific generators). Because the parsed/annotated AST is shared across generators, behavioral semantics are bound a single time at the adaptor boundary and reused end-to-end—turning the classic N×M mapping problem into N+M (N sources adapted to M, M targets generated from M). This positions M as a reusable, semantics-aware pivot that eases simulation of multiple DSLs and code generation toward multiple platforms without duplicating low-level mapping logic in each toolchain branch. 

In short, metamodel-based DSLs are ideal for defining domains, while M is designed for executing them: by adapting DSLs to M, organizations gain a single, formalized gateway to simulation, verification, and portable implementation with improved semantic consistency across artifacts.
\subsection{M vs. Modelica}
Modelica is a top candidate we initially thought of for the M language as it is a well-designed, popular textual object-oriented modeling and simulation language that supports both continuous and discrete systems. However following considerations hinder us from selecting Modelica as M:
\begin{itemize}
    \item Concurrency model: Modelica still has to rely on multi-threaded synchronization mechanisms (like Mutex) for handling concurrency~\cite{modelica_concurrency}. In our opinion, the actor model is more simple, disciplined and scalable, while the concurrency model with threads and shared memory is prone to bugs such as data races, deadlocks, and atomicity violations \cite{DBLP:conf/birthday/AghaP18}.
    \item Messaging semantics: The equation-based connection in Modelica is more suitable for modeling physical connections like electric wiring, fluid pipes. The messaging semantics is different and needs to be added as a workaround \cite{pdevsmodelica,DBLP:journals/simulation/SanzU23}.
    \item Equation solving: the differential equation system in Modelica has to be flattened and solved by the compiler while the compiling results are not made available for reusing~\cite{DBLP:books/daglib/0020345}. This implies redoing compilation and complications in transformation and implementation of such in the other models.
\end{itemize}
Currently, M models continuous behaviours by stepwise discretization. Straightforward continuous modeling and simulation through differential equations and numerical integrators is planned in a future project as we have to consider equivalent expressiveness, realizability and verifiability in different languages. For current phase, we accept this shortcoming of M to be able to move forward with the toolchain demonstration. 

\subsection{M vs. Ptolemy and Lingua Franca}
Ptolemy II \cite{Ptolemaeus:14:SystemDesign}\footnote{\url{https://ptolemy.berkeley.edu/}} and Lingua Franca (LF) \cite{10.1145/3617687} are syster projects led by Professor Edward Lee of UC Berkeley. Both are based on the actor model for modeling, simulation, and implementation of concurrent, real-time, embedded systems. Ptolemy II  is a mature component-oriented graphical modeling and design environment with actor-based components that can be coupled together using in/out ports. Behind the scene, Ptolemy II uses an XML-based language (named MoML) to describe the model and allows adding new actors with customized code in Java (\textit{which makes the semantics of the model non-declarative and buried among implementation details}). Ptolemy II supports heterogeneous concurrent models of computation (MoCs). Rather than committing to a single concurrency paradigm, Ptolemy II allows combining dataflow, DE \cite[ch.~7]{Ptolemaeus:14:SystemDesign}, synchronous-reactive, continuous-time, and process networks within the same system model. This makes it highly flexible for multi-domain modeling, but at the cost of a more complex semantic framework compared to M’s focus on a single formalism. In comparison, M provides a streamlined, DEVS-inspired execution model where time and causality are central, while Ptolemy II focuses on heterogeneity and tool integration, offering a broad experimentation platform for exploring different MoCs. Thus, M emphasizes formal simplicity and precision in execution semantics, whereas Ptolemy II emphasizes flexibility and extensibility for modeling diverse systems.

LF is a coordinating language for specifying and implementing concurrent, distributed, embedded systems more dependably. LF has hosting mechanisms to make actual execution time align with logical time specification. As a textual actor-based language, LF is also a top candidate for us to consider. However, the features below are barriers for LF to be M:
\begin{itemize}
    \item Interleaving of two timelines: mixing logical timeline and physical timeline in a model will make it harder to interpret accordingly in other models.
    \item Interleaving of modeling language and implementation language: LF allows inline programming in an external language like C, Python, Rust to finally generate implementation code in that language. It is good for implementing purpose but makes it harder to understand the full semantics of the model and convert to other formalisms.
    \item Forced determinism: the determinism of LF programs dedicately relies  on their own execution platform, which may not be achieved in other execution platforms (e.g. ROS2).
    \item Currently LF does not provide declaration of state-based events and mechanism for scanning state changes and firing matched events.
\end{itemize}
Instead we will later investigate if we can use LF as a code generator for M, as verification of semantic equivalence between M and LF is needed.

\section{Current Status and Future Plans} \label{sec:progress}
The initial design phase of M has been completed, including the specification of its syntax in ANTLR4 and its abstract syntax in JastAdd. At this stage, the compiler is capable of parsing M programs, constructing and decorating the AST, and performing preliminary code generation. Ongoing work addresses semantic analyses to ensure contextual soundness. As language and compiler development is inherently iterative and time-consuming, our immediate objective is to establish a prototype that demonstrates the feasibility of the toolchain. Once stabilized, the framework will be experimented on representative case studies with model-driven engineering tasks around M: modeling, simulation, deployment, and formal verification. The whole solution will be released as open source in a subsequent stage of the project.

\section{Conclusion} \label{sec:conclude}
Modern systems continue to grow in complexity, and model-based engineering offers a means to abstract away from low-level implementation concerns toward higher-level conceptual models. This shift promises faster development, improved predictability, and increased reliability in system design. With M, we aim to advance the state of the art in model-driven system and software engineering by placing software language engineering at the heart. Once the toolchain is fully established, adapters can be developed to seamlessly map existing modeling languages into M, thereby enabling semantics-aware model compilation with minimal additional effort. The M toolchain is envisioned as an end-to-end ecosystem that unifies modeling, simulation, execution, and verification. It effectively shields system developers from the intricacies of underlying formal mechanisms, allowing them to focus on design tasks while relying on M’s semantics and tooling. In this way, M not only strengthens methodological rigor but also broadens accessibility and applicability in model-based engineering practice.

%Systems are getting more complex, and model-based engineering is a way of escaping from low-level details to focus on higher-level views, deemed to help design and create systems faster, more predictable, more reliable. Our work will advance the state-of-art in model-based systems and software engineering, shifting the approach more towards software language engineering. Once the toolchain is in place, adaptors can be developed to map other modeling languages to M and a semantics-aware model compilation paradigm is ready with little or no extra custom work. The M toolchain forms a full end-to-end ecosystem of engineering tools including modeling, simulation, execution, verification without requiring the system developers to know and implement complicated mechanisms behind them other than adapting to or using the M language directly.

\section{Acknowledgment}
%.%
%[Hidden due to double-blind review policy] 
%\begin{anonsuppress}
This work is funded by the Swedish Research Council through the ORPHEUS project (rn. 2022-03408).
%We thank XYZ for proof-reading and giving meaningful comments to help improve the presentation.
%\end{anonsuppress}

%\bibliographystyle{ACM-Reference-Format}
%\nocite{*}
\bibliography{m-arxiv}

\end{document}